\documentclass[notitlepage,aps,prd,superscriptaddress,preprintnumbers,nofootinbib,longbibliography,showpacs]{revtex4-1}
\usepackage[%
  colorlinks=true,
  urlcolor=blue,
  linkcolor=blue,
  citecolor=blue
]{hyperref}
\usepackage{slashed,color,amsmath,amssymb,mathrsfs}
\usepackage{graphicx}
\usepackage{hyperref}
\usepackage{longtable}
\usepackage{array}

\allowdisplaybreaks

\def\gamf{\gamma_5}
\def\la{\langle}
\def\ra{\rangle}

\def\chip{\chi_+}

\def\chim{\chi_-}

\def\fp{f_+}
\def\fm{f_-}

\def\chid{{\chi^\dag}}

\def\ud{{u^\dag}}

\def\Gam{\Gamma}

\def\fgamma{\gamf\gamma}

\def\Bb{\bar{B}}
\def\B{B}

\def\As{A}
\def\Ase{A}
\def\Ae{A}

\begin{document}
\title{Meson-baryon effective chiral Lagrangians at order $p^4$}
\author{Shao-Zhou Jiang}
\email[]{jsz@gxu.edu.cn}
\affiliation{Department of Physics and GXU-NAOC Center for Astrophysics and Space Sciences, Guangxi University,
Nanning, Guangxi 530004, People's Republic of China}
\affiliation{Guangxi Key Laboratory for the Relativistic Astrophysics, Nanning, Guangxi 530004, People's Republic of China}
\author{Qing-Sen Chen}
\email[]{chenqs16@mails.jlu.edu.cn}
\affiliation{Department of Physics and GXU-NAOC Center for Astrophysics and Space Sciences, Guangxi University,
Nanning, Guangxi 530004, People's Republic of China}

\author{Yan-Rui Liu}
\email{yrliu@sdu.edu.cn}
\affiliation{School of Physics and Key
Laboratory of Particle Physics and Particle Irradiation (MOE),
Shandong University, Jinan 250100, China}
\begin{abstract}
We construct the three-flavor Lorentz-invariant meson-baryon chiral Lagrangians at the order $p^4$, with which a full one-loop investigation may be performed. One obtains 540 independent terms. The processes with the minimal number of mesons and photons that this order Lagrangians may contribute to are also presented.
\end{abstract}
 \maketitle
\section{Introduction}\label{intr}

It is difficult for us to deal with strong interactions with the underlying QCD in the low energy region because of the nonperturbative effects. Thanks to the chiral symmetry of QCD and its spontaneous breaking, we may equivalently describe low-energy physics involving pseudoscalar pions with an effective theory at hadron level, chiral perturbation theory (ChPT) \cite{weinberg,GS1,GS2,Gasser:1987rb}. The perturbative expansion is organized with the pion momentum ($p$) and the description avoids the complex interactions between quarks and gluons. In this effective field theory, the Lagrangian contains all chirally invariant terms. According to Weinberg's power counting rules, the needed number of terms is finite when one calculates the $T$-matrix of a process to the required order, because the theory is renormalised order by order. At present, the chiral Lagrangians (CLs) of pseudoscalar mesons have been constructed up to the order ${\mathcal O}(p^6)$ (two-loop level) for both normal and anomalous parts \cite{GS1,GS2,p61,p62,p6p,p6a1,p6a2,tensor1,U3,ourf}. They contain the whole 16 bilinear light-quark currents (scalar, pseudoscalar, vector, axial-vector, and tensor) of the special unitary group and the unitary group.

Matter fields (baryons, heavy mesons, etc.) can be introduced into the framework of chiral perturbation theory with the transformation of $SU(2)_V$ or $SU(3)_V$. For the pion-nucleon interaction, the relevant CLs have been obtained up to the order ${\mathcal O}(p^4)$ (one-loop level) \cite{Krause:1990xc,Ecker:1994pi,1998NuPhA.640..199F,Meissner:1998rw,pin4}. To study the pion-hyperon interaction in a model-independent approach, one needs the $SU(3)$ meson-baryon CLs. At present, the terms at the order ${\mathcal O}(p^3)$ have been obtained \cite{pib31,pib32}. However, the full one-loop level investigation needs next-order terms.

The inclusion of nucleons in the chiral perturbation theory makes the chiral expansion of $p/\Lambda_\chi$ problematic because both the baryon mass $m_N$ and the scale of chiral symmetry breaking $\Lambda_\chi$ are around 1 GeV. This problem is cured in the heavy baryon chiral perturbation theory (HBChPT), where the large parameter $m_N$ is eliminated and an exact power counting rule exists. Although the expansion is convergent and the theory works well in the pion-nucleon system \cite{Bernard:1992qa,Ecker:1995rk,1998NuPhA.640..199F,pin3b}, the convergence in the kaon-hyperon systems is slow or violated \cite{Liu:2006xja}. In the three-flavor meson-baryon scattering processes at threshold, several Weinberg-Tomozawa terms vanish and the ${\cal O}(p^4)$ calculation may answer whether the heavy baryon formalism still works or not in these channels. To recover the relativistic formulation of the theory, the infrared regularization scheme \cite{Ellis:1997kc,Becher:1999he} and the extended on-mass-shell (EOMS) renormalization scheme \cite{Fuchs:2003qc} are proposed. The latter scheme seems to work well \cite{Kubis:2000zd,Kubis:2000aa,Zhu:2000zf,Zhu:2002tn,Frink:2004ic,Schindler:2003xv,Schindler:2005ke,Geng:2008mf,Alarcon:2011kh,Alarcon:2011zs,Alarcon:2012kn,Chen:2012nx,Sun:2016wzh}. Since the $SU(3)$ baryon ChPT involves the larger kaon mass, further explorations with the ${\cal O}(p^4)$ Lagrangians are helpful to understand the convergence of the chiral expansion.

In baryon ChPT, not all interaction terms at the same order contribute to the $T$-matrix of a special process. There are studies involving high order interactions in the literature although the complete Lagrangian is not known. In Refs. \cite{nmp50,nmp5,nmp61,nmp62}, the nucleon masses were calculated up to the order ${\cal O}(p^6)$ (two-loop level). In Ref.  \cite{bmp4}, the ground-state octet baryon masses and sigma terms were studied up to the order ${\cal O}(p^4)$ where seven ${\cal O}(p^4)$ terms are involved. The electromagnetic form factors of the ground state baryon octet were also studied up to the order ${\cal O}(p^4)$ with five terms at that order \cite{bemffp4}. Needless to say, the complete ${\cal O}(p^4)$ chiral Lagrangians which are constructed systematically are helpful for further studies of meson-baryon processes.

An obvious obstacle in the application of ChPT is the determination of the parameters in the Lagrangian, the so-called low energy constants (LECs). In principle, once the parameters are determined up to some chiral orders, they are applicable to all processes to those orders. Because of the difficulty in low energy QCD, one usually extracts the LECs by fitting experimental data, a feasible method to determine them. It seems that we still need more meson-baryon scattering data and high order Lagrangians are not useful. However, efforts have been tried to determine all the LECs from QCD in the meson sector \cite{Jiang:2009uf,Jiang:2010wa,Jiang:2012ir,Jiang:2015dba} and similar study in the baryon sector is on the way. Various models and symmetries are also used to constrain the values of LECs or relations between them \cite{Bernard:1996gq,Luo:2006yg,Sanz-Cillero:2013ipa,Du:2016tgp}. These studies make further investigations with the baryon ChPT possible although the number of LECs becomes large for high order Lagrangians. With the development of computing capacity, the lattice simulation or other numerical calculations will be helpful to fix the LECs. Now, the investigations with ChPT in the meson sector at two loops are relatively easy to perform once the LECs are known \cite{Bijnens:2014gsa}. In the baryon sector, similar investigations are also possible. The starting point for such studies is, of course, the meson-baryon chiral Lagrangian.

In this paper, we would like to complete the Lagrangians of the $SU(3)$ meson-baryon chiral perturbation theory to
the one-loop ${\cal O}(p^4)$ order. Deeper understanding on the low-energy meson-baryon interactions needs these terms. This work is organized as follows: In Sec. \ref{rev}, we review the building blocks for the construction of the chiral Lagrangians. In Sec. \ref{const}, a systematic method for the construction is introduced, which is based on the properties of the building blocks and the linear relations to the final results. In Sec. \ref{results}, we list our results and give a discussion. Section \ref{summ} is a short summary.

\section{Building blocks of chiral Lagrangians}\label{rev}
The QCD Lagrangian $\mathscr{L}$ can be written as
\begin{align}
\mathscr{L}=\mathscr{L}^0_{\mathrm{QCD}}
+\bar{q}(\slashed{v}+\slashed{a}\gamma_5-s+ip\gamma_5)q~,
\end{align}
where $\mathscr{L}^0_{\mathrm{QCD}}$ is the original QCD lagrangian, $q$ denotes the quark field, and $s$, $p$, $v^\mu$, and $a^\mu$ denote scalar, pseudoscalar, vector, and axial-vector external sources, respectively. Conventionally, we ignore the tensor source and the $\theta$ term here.

If the light quarks are massless, the QCD Lagrangian $\mathscr{L}^0_{\mathrm{QCD}}$ exhibits a global $SU(3)_L\times SU(3)_R$ chiral symmetry. In the description of the effective chiral Lagrangiang, the Goldstone bosons (pseudoscalar mesons) coming from the spontaneous chiral symmetry breaking are collected into an $SU(3)$ matrix $U$. If the chiral rotation for the light quarks is $g=(g_L, g_R)$, the transformation for the meson field would be $U\to g_L U g_R^\dag$. It is convenient to introduce $u^2=U$ to simplify the construction of the chiral Lagrangians. The field $u$ transforms as $u\to g_L uh^\dag=h ug_R^\dag$ under the chiral rotation, where $h$ in $SU(3)_V$ from the breaking of $SU(3)_L\times SU(3)_R$ is a function of the pion fields. The baryon octet is denoted by another matrix $B$,
\begin{align}
B=\begin{pmatrix}
\frac{\Sigma^0}{\sqrt{2}}+\frac{\Lambda}{\sqrt{6}} & \Sigma^+ & p\\
\Sigma^- & -\frac{\Sigma^0}{\sqrt{2}}+\frac{\Lambda}{\sqrt{6}} & n\\
\Xi^- & \Xi^0 & -\frac{2\Lambda}{\sqrt{6}}
\end{pmatrix}.
\end{align}
Under the chiral rotation, this matrix transforms into $B'=hBh^\dag$.

To construct the Lagrangians, one collects the external sources and meson and baryon fields together and define appropriate combinations of them. The fields or combinations transforming like the baryon field are called building blocks. The building blocks we will use are
\begin{align}
u^\mu&=i\{u^\dag(\partial^\mu-ir^\mu)u-u(\partial^\mu-il^\mu)u^\dag\},\notag\\
\chi_\pm&=u^\dag\chi u^\dag\pm u\chi^\dag u,\notag\\
h^{\mu\nu}&=\nabla^\mu u^\nu+\nabla^\nu u^\mu,\notag\\
f_+^{\mu\nu}&=u F_L^{\mu\nu} u^\dag+ u^\dag F_R^{\mu\nu} u,\notag\\
f_-^{\mu\nu}&=u F_L^{\mu\nu} u^\dag- u^\dag F_R^{\mu\nu} u=-\nabla^\mu u^\nu+\nabla^\nu u^\mu,\notag\\
B,\Bb,&\label{df}
\end{align}
where $r^\mu=v^\mu+a^\mu$, $l^\mu=v^\mu-a^\mu$, $\chi=2B_0(s+ip)$, $F_R^{\mu\nu}=\partial^{\mu}r^{\nu}-\partial^{\nu}r^{\mu}-i[r^{\mu},r^{\nu}]$, $F_L^{\mu\nu}=\partial^{\mu}l^{\nu}-\partial^{\nu}l^{\mu}-i[l^{\mu},l^{\nu}]$ and $B_0$ is a constant related to the quark condensate. In this paper, we consider the case in which both $v^\mu$ and $a^\mu$ are traceless. The definition of the covariant derivative $\nabla^{\mu}$ acting on any building block $O$ is
\begin{align}
\nabla^{\mu}O&=\partial^\mu O+[\Gamma^\mu,O],\label{cd}\\
\Gamma^{\mu}&=\frac{1}{2}\{\ud(\partial^\mu-ir^\mu)u+u(\partial^\mu-il^{\mu})\ud\}.\label{cd1}
\end{align}
With this definition, all building blocks $O$'s, including their covariant derivative forms, transform into $O'=hOh^\dag$ under the chiral rotation. As in Ref. \cite{p62}, we also define $\chi_{\pm}^{\mu}$ to substitute the covariant derivative of $\chi_\pm$,
\begin{align}
\chi_{\pm}^{\mu}=u^\dag \widetilde{\nabla}^\mu \chi u^\dag\pm u \widetilde{\nabla}^\mu\chid u=\nabla^\mu\chi_{\pm}-\frac{i}{2}\{\chi_{\mp},u^\mu\},\label{chimu}
\end{align}
because sometimes it is more convenient. Here, $\widetilde{\nabla}^\mu\chi\equiv \partial^\mu\chi-ir^\mu\chi+i\chi l^\mu$ \footnote{The details of the covariant derivatives of $\chi$, $\chi^\dag$ and $F_{L,R}^{\mu\nu}$ can be found in Ref. \cite{p61}.}.
In addition, the following two relations of covariant derivatives are useful,
\begin{align}
&[\nabla^\mu,\nabla^\nu]O=[\Gamma^{\mu\nu},O],\label{Gam0}\\
&\Gamma^{\mu\nu}=\nabla^{\mu}\Gam^{\nu}-\nabla^{\nu}\Gam^{\mu}-[\Gam^{\mu},\Gam^{\nu}]
=\frac{1}{4}[u^\mu,u^\nu]-\frac{i}{2}f_{+}^{\mu\nu}.\label{Gam1}
\end{align}

\section{Construction of chiral Lagrangians}\label{const}
With the above building blocks, one can construct the SU(3) meson-baryon chiral Lagrangians following the steps given in this section. The final Lagrangian contains minimal independent chirally invariant terms. This method is very similar to the construction of the meson chiral Lagrangians in Ref. \cite{ourf}.

\subsection{Power counting and transformation properties}
The construction of chirally invariant monomials needs the power counting of the building blocks. We show their chiral dimensions in the second column of Table \ref{blbt} \cite{GS1,GS2,Gasser:1987rb,p62,pin4,pib31}. The covariant derivative acting on the meson fields or external sources is counted as ${\mathcal O}(p^1)$, but that acting on the baryon fields is counted as ${\mathcal O}(p^0)$. For convenience, we will use $D^\mu$ rather than $\nabla^\mu$ to denote the covariant derivative in the latter case. There is no difference between $\nabla^\mu$ and $D^\mu$. According to the forms of the bilinear couplings of $\Bb$ and $B$ in the low-energy approximation, one assigns the chiral dimensions of the elements of the Clifford algebra basis in the second column of Table \ref{cabt} \cite{pin4,pib31,Scherer:2012xha}.
\begin{table*}[!h]
\caption{\label{blbt}Chiral dimension (Dim), parity ($P$), charge conjugation ($C$) and hermiticity of the building blocks.}
\begin{tabular}{ccccc}
	\hline\hline
	                   & Dim &         $P$         &            $C$             &        h.c.         \\ \hline
	    $u^{\mu}$      & 1 &     $-u_{\mu}$      &       $(u^{\mu})^T$        &      $u^{\mu}$      \\
	   $h^{\mu\nu}$    & 2 &    $-h_{\mu\nu}$    &      $(h^{\mu\nu})^T$      &    $h^{\mu\nu}$     \\
	   $\chi_{\pm}$    & 2 &   $\pm\chi_{\pm}$   &      $(\chi_{\pm})^T$      &  $\pm \chi_{\pm}$   \\
	$f_{\pm}^{\mu\nu}$ & 2 & $\pm f_{\pm\mu\nu}$ & $\mp (f_{\pm}^{\mu\nu})^T$ & $ f_{\pm}^{\mu\nu}$ \\
	       $B$         & 0 &         $B$         &           $B^T$            &         $B$         \\
	      $\Bb$        & 0 &        $\Bb$        &          $\Bb^T$           &        $\Bb$        \\
	    $D^\mu B$      & 0 &      $D_\mu B$      &       $-(D^\mu B)^T$       &     $-D^\mu B$      \\ \hline\hline
\end{tabular}
\end{table*}

\begin{table*}[!h]
\caption{\label{cabt}Chiral dimension (Dim), parity ($P$), charge conjugation ($C$) and hermiticity of the Clifford algebra elements.}
\begin{tabular}{ccccc}
	\hline\hline
	                  & Dim & $P$ & $C$ & h.c. \\ \hline
	       $1$        & 0 & $+$ & $+$ & $+$  \\
	     $\gamf$      & 1 & $-$ & $+$ & $-$  \\
	 $\gamma^{\mu}$   & 0 & $+$ & $-$ & $+$  \\
	 $\fgamma^{\mu}$  & 0 & $-$ & $+$ & $+$  \\
	$\sigma^{\mu\nu}$ & 0 & $+$ & $-$ & $+$  \\ \hline\hline
\end{tabular}
\end{table*}

The chiral Lagrangian is also invariant under the transformation of parity ($P$), charge conjugation ($C$) and Hermitian conjugation (h.c.). The transformation properties of the building blocks are simple \cite{p62,pin4,pib31} and we collect them in Table \ref{blbt}, where the minus sign for $D^\mu B$ is from the moving of the derivative (see Eq. \eqref{partialintegration}). However, the properties of the Clifford algebra are slightly complicated. We here adopt the method used in Ref. \cite{pin4} to analyze the transformations. Generally speaking, the invariant monomials have the following forms,
\begin{align}
\la\Bb A^{\mu\nu\cdots}\Theta_{\mu\nu\cdots}B\ra+\mathrm{h.c.},\hspace{2ex}
\la\Bb A_1^{\mu\cdots}\ra\la A_2^{\nu\cdots}\Theta_{\mu\nu\cdots}B\ra+\mathrm{h.c.},\hspace{2ex}\cdots,\label{clform}
\end{align}
where $\la\cdots\ra$ is a trace over $SU(3)$ indices, $A^{\mu\nu\cdots}$, $A_1^{\mu\cdots}$ and $A_2^{\nu\cdots}$ are the products of meson fields and/or external sources, and $\Theta_{\mu\nu\cdots}$ is the product of a Clifford algebra element $\Gamma\in\{1,\gamma_\mu,\gamf,\fgamma_\mu,\sigma_{\mu\nu}\}$, the Levi-Civita tensor $\varepsilon^{\alpha\beta\rho\tau}$, and several covariant derivatives $D_{\lambda}D_{\eta}\cdots$ acting on $B$. The forms in Eq. \eqref{clform} can have other deformations, such as increasing traces, a change of the place of the $B$ field, and so on. For the $P$ transformation, the $\gamma$ matrices are changed by raising or lowering the Lorentz indices. Table \ref{cabt} only lists the extra signs. For the $C$ and h.c. transformations, one may change $\la\Bb A^{\mu\nu\cdots}\Gamma_{\mu\cdots}D_{\nu\cdots}B\ra$ to $\pm\la D_{\nu\cdots}\Bb A^{\prime\mu\nu\cdots}\Gamma_{\mu\cdots}B\ra$ with an extra transpose or a whole hermitian conjugation,  where $A^{\prime\mu\nu\cdots}$ comes from $A^{\mu\nu\cdots}$. Table \ref{cabt} lists the plus or minus sign. Other forms of monomials can be operated in a similar way. Then, it is easy to move the covariant derivatives acting on $\Bb$ to $B$ with the partial integration relation (Eq. (\ref{partialintegration})). The signs in both Table \ref{blbt} and Table \ref{cabt} are given with this consideration in mind.

\subsection{Linear relations}\label{lr}
In general, the invariant monomials that are combined with the building blocks are not independent. Several linear relations exist that can be used to find the independent monomials. We collect these relations as follows.
\begin{enumerate}
\item \label{pirl}Partial integration.

The covariant derivative acting on the whole monomial can be discarded and one has
\begin{align}
0=&\la (\nabla^\mu A) B\cdots\ra\la CD\cdots\ra\cdots+\la A (\nabla^\mu B)\cdots\ra\la CD\cdots\ra\cdots\notag\\
&+\la A B\cdots\ra\la (\nabla^\mu C)D\cdots\ra\cdots +\la A B\cdots\ra\la C(\nabla^\mu D)\cdots\ra\cdots+\text{other terms},\label{pir}
\end{align}
where ``$\cdots$" represents one or more building blocks. Because the covariant derivative acting on baryon fields is counted as ${\mathcal O}(p^0)$ and that on other building blocks is ${\mathcal O}(p^1)$, we can simply employ the relation \cite{pin4,pib31}
\begin{align}\label{partialintegration}
\la D_{\nu}\Bb A^{\mu\nu\cdots}\Gamma_{\mu\cdots}B\ra\doteq-\la\Bb A^{\mu\nu\cdots}\Gamma_{\mu\cdots}D_{\nu}B\ra
\end{align}
in reducing the number of monomials. The symbol ``$\doteq$" means that both sides are equal if high order terms are ignored. This is also the origin of the extra minus signs in Table \ref{blbt} and Table \ref{cabt} discussed above.

\item Equations of motion (EOM).

The lowest order EOM from the pseudoscalar CL is
\begin{align}
\nabla_\mu u^\mu&=\frac{i}{2}\bigg(\chim-\frac{1}{N_f}\la\chim\ra\bigg),\label{eomb}
\end{align}
where $N_f$ is the number of quark flavors and we take $N_f=3$ here. This equation indicates that the monomials including $\nabla_\mu u^\mu$ can be removed. Obviously, the higher order EOM only adds terms on the right hand side and it has no impact on the construction of CL. The EOM from the meson-baryon CL is a little complicated, which limits the forms of $\Theta_{\mu\nu\cdots}$ to a small set. We leave the discussions of this for Appendix \ref{aeom}.

\item Bianchi identity.

From Eqs. \eqref{Gam0} and \eqref{Gam1}, one gets
\begin{eqnarray}
\nabla^\mu\Gamma^{\nu\lambda}+\nabla^\nu\Gamma^{\lambda\mu}+\nabla^\lambda\Gamma^{\mu\nu}=0,\label{bi}
\end{eqnarray}
which gives a relation between the covariant derivatives of $\Gamma^{\mu\nu}$ (or $\fp^{\mu\nu}$).

\item Schouten identity.

When $\epsilon^{\mu\nu\lambda\rho}$ exists in a monomial, the Schouten identity indicates that
\begin{eqnarray}
\epsilon^{\mu\nu\lambda\rho}A^\sigma-\epsilon^{\sigma\nu\lambda\rho}A^\mu-\epsilon^{\mu\sigma\lambda\rho}A^\nu
-\epsilon^{\mu\nu\sigma\rho}A^\lambda-\epsilon^{\mu\nu\lambda\sigma}A^\rho=0.\label{sir}
\end{eqnarray}

\item Cayley-Hamilton relation.

All the building blocks are $3\times3$ matrices in flavor space. For any $3\times3$ matrices $A$, $B$ and $C$, one has
\begin{eqnarray}
0&=&ABC+ACB+BAC+BCA+CAB+CBA-AB\la C\ra-AC\la B\ra-BA\la C\ra-BC\la A\ra-CA\la B\ra\notag\\
&&-CB\la A\ra-A\la BC\ra-B\la AC\ra-C\la AB\ra-\la ABC\ra-\la ACB\ra+A\la B\ra\la C\ra+B\la A\ra\la C\ra+C\la A\ra\la B\ra\notag\\
&&+\la A\ra\la BC\ra+\la B\ra\la AC\ra+\la C\ra\la AB\ra-\la A\ra\la B\ra\la C\ra,\label{ch3}
\end{eqnarray}
which is the Cayley-Hamilton relation \cite{p62,ourf}.

\item Contact terms.

In constructing the CL, we need to consider separately the contact terms where only baryon fields and pure external sources ($F_R^{\mu\nu}$, $F_L^{\mu\nu}$, and $\chi$) are involved. To adopt the constraint relations, we also use the following formulas by revealing explicitly the sources in Eq. \eqref{df},
\begin{align}
F_L^{\mu\nu}&=\frac{1}{2}u^\dag(f_+^{\mu\nu}+f_-^{\mu\nu})u,\notag\\
F_R^{\mu\nu}&=\frac{1}{2}u(f_+^{\mu\nu}-f_-^{\mu\nu})u^\dag,\notag\\
\chi&=\frac{1}{2}u(\chi_++\chi_-)u,\notag\\
\chi^\dag&=\frac{1}{2}u^\dag(\chi_+-\chi_-)u^\dag.\label{lrct}
\end{align}
The number of such terms is small and it is not difficult to construct them directly. The ${\mathcal O}(p^4)$ meson-baryon contact terms are the last three monomials in Table \ref{p4rtab}.
\end{enumerate}

\subsection{Reduction of the monomials}\label{redm}
Because the number of $D^\mu$ acting on $B$ is arbitrary, it seems that there are infinite possibilities of monomials at a given order. However, from Eq. \eqref{Gam0} and the item (\ref{dd}) in Appendix \ref{aeom}, one finds that the covariant derivatives $D_{\nu}D_{\lambda}\cdots$ acting on $B$ have totally symmetric Lorentz indices and any two Lorentz indices are completely different. To reflect the symmetric nature, one may use the short notation $D_{\nu\lambda\rho\cdots}$ to denote multiple derivatives where
\begin{align}
D_{\nu\lambda\rho\cdots}=D_\nu D_\lambda D_\rho\cdots+ \text{full permutation of $D$'s.}
\end{align}
This symmetric property limits the possibilities of the monomials.

On the other hand, some monomials with a different order of building blocks and different indices may be equal. The construction of independent monomials will be easier if we change all the monomials to a unified form. The following rules are helpful.
\begin{enumerate}
\item Unlike the two-flavor $\pi N$ CL in Ref. \cite{pin4}, where the nucleon field $\bar{\Psi}$ is fixed on the far left and the $\Psi$ is fixed on the far right, now the positions of the baryon fields $\Bb$ and $B$ are not fixed except that $B$ is always on the right hand of $\Bb$. To fix the positions of $\Bb$ and $B$, we first move the trace containing $\Bb$ to the left, and then move the field $\Bb$ to the far left. If $B$ is in another trace, we move the trace to the right hand side of the trace containing $\Bb$, and then move $\B$ to the far right inside the trace. With this rule, the positions of $\Bb$ and $B$ are fixed. There may also exist traces containing neither $\Bb$ nor $B$. We move them to the right hand side of the trace containing $B$. The relative positions of these traces are also not fixed. We treat them in the next item. The fixed form of the monomials is like that in Eq. \eqref{clform} where the factor $\Theta_{\mu\nu\cdots}$ is moved to the left side of $B$.

\item For a trace without $\Bb$ and $B$, another rule is introduced. All building blocks are numbered, including the covariant derivative $\nabla$. Table \ref{noo} shows an example. The meaning of the number for each building block is not significant. What we care about is the relative size. Each cyclic permutation maps to a vector, such as $\la u h\ra\to (1161, 1181)$, $\la h u\ra\to (1181, 1161)$, $\la \chip\ra\to (1201)$ and $\la \chim\ra\to (1221)$. We choose the smallest permutation so that the smaller number is placed as far left as possible. For example, $(1161, 1181)$ is smaller than $(1181, 1161)$ and thus we choose the combination $\la u h\ra$ but not $\la h u\ra$. If there is more than one trace without $\Bb$ and $B$, we place the smaller one on the left. For example, $(1201)$ is smaller than $(1221)$, so we choose $\la \chip\ra\la\chim\ra$ but not $\la\chim\ra\la\chip\ra$.

\begin{table*}[h]
\caption{\label{noo}Examples of numbering for the building blocks (or operator) and the Lorentz indices. No significant meaning is given to the numbers but the relative size of the numbers is meaningful.}
\begin{tabular}{rccccccccccccc}
	\hline\hline
	Operator & $\nabla$ &  $u$  &   $\fm$   &  $h$   &  $\fp$   & $\chip$ & $\chim$  \\
	\hline
	  Number &   1101   & 1161  &   1171    &  1181  &   1121   &  1201   &  1221   \\
	\hline\hline
	   Index &  $\mu$   & $\nu$ & $\lambda$ & $\rho$ & $\sigma$   \\
	\hline
	  Number &    1     &   2   &     3     &   4    &    5       \\
	\hline\hline
\end{tabular}
\end{table*}

\item For the Lorentz indices, the rule is the same as the building blocks. All indices are numbered, too. We also give an example in Table \ref{noo}. When the places of all the building blocks are fixed, their indices are mapped to  vectors, such as $\la u^{\mu}u^{\nu}u^{\mu}\ra\to(1,2,1)$, $\la u^{\mu}u^{\mu}u^{\nu}\ra\to(1,1,2)$ and $\la u^{\nu}u^{\lambda}u^{\lambda}\ra\to(2,3,3)$, and so on. Although the results are probably equal (as shown here), we only choose the smallest permutation, $\la u^{\mu}u^{\mu}u^{\nu}\ra\to(1,1,2)$. In this step, the Einstein summation convention and the symmetric and antisymmetric relations for $f^{\mu\nu}_\pm$, $h^{\mu\nu}$ and $\epsilon^{\mu\nu\lambda\rho}$ are used.
\end{enumerate}

We say that a monomial obeying the above rules has a {\it standard form}. With these rules, two monomials having the same standard form are equal. The final results are all in this form. Besides the purpose of distinguishing monomials, the standard form is also conveniently used for programming.

\subsection{Classifications and Substitutions}
It is not complicated to obtain all possible invariant monomials at a given order with the building blocks $u^\mu$, $h^{\mu\nu}$, $\chi_\pm$, $f^{\mu\nu}_\pm$, and their derivative forms and the constrained $\Theta_{\mu\nu\cdots}$ in Appendix \ref{aeom}. However, the number of the resulting monomials is too large and it makes further manipulation difficult. A simpler way is to classify all the monomials according to the external sources. It means that we can treat first the category with four pseudoscalar sources, then the category with three pseudoscalar sources plus one vector current (or one covariant derivative, see Eq. \eqref{cd}), and so on. One may adopt such a classification because almost all the linear relations in the subsection \ref{lr} connect monomials with the same type of external sources and one applies those relations to monomials category by category. The exceptional case is for the contact terms where different types of external sources may be connected with the relations in the subsection \ref{lr}. We will deal with this case separately.

To simplify the calculation, we usually make the following replacements,
\begin{align}
\fp^{\mu\nu}\longleftrightarrow i\Gamma^{\mu\nu},\hspace{0.2cm}
\chi_{\pm}^\mu\longleftrightarrow\nabla^\mu\chi_\pm.\label{rdb}
\end{align}
Since our purpose is to construct all the ${\mathcal O}(p^4)$ CL, the differences induced by these replacements can be compensated by other terms at the same chiral order. That is, in constructing the CL, we use actually the definitions $\chi_{\pm}^\mu=\nabla^\mu\chi_\pm$ and $\Gamma^{\mu\nu}=-i\fp^{\mu\nu}$ rather than the strict ones in Eqs. \eqref{chimu} and \eqref{Gam1}.

\subsection{Independent linear relations and chiral Lagrangians}

With the above preparations, now we can move on to find out independent chiral-invariant terms with a systematic approach easy to program. This approach has been used to construct meson chiral Lagrangians in Ref. \cite{ourf}.

First, one sets up basic equations. So we may adopt the linear relations in the subsection \ref{lr} directly, it is convenient for us to reveal the covariant derivatives in the constructed monomials by using Eqs. \eqref{df} and \eqref{Gam1}. Here, we use $D_{i,j}$ to store all possible invariant monomials constructed with $\bar{B}$, $B$, $u^\mu$, $\chi_\pm,h^{\mu\nu}$, $\Gamma^{\mu\nu}$, $f_-^{\mu\nu}$, and their derivative forms and $E_{i,j}$ to store all possible monomials revealing the covariant derivatives (constructed with $\bar{B}$, $B$, $u^\mu$, $\chi_\pm$, $\Gamma^{\mu}$, and their derivative forms). The index $i$ labels the categories and the index $j$ labels the monomials inside the category $i$. The linear relations between $D_{i,j}$ and $E_{i,j}$ are
\begin{align}
D_{i,j}=\sum_{k}A_{i,jk}E_{i,k},\label{lr0}
\end{align}
where the coefficient matrix $A_{i}$ for the category $i$ is easy to obtain with Eqs. \eqref{df} and \eqref{Gam1}.

Second, one finds out independent constraint relations. By applying the linear relations in the subsection \ref{lr} to $E_{i,k}$, we obtain the constraint equations
\begin{align}
\sum_{k}R_{i,jk}E_{i,k}=0,\label{rls}
\end{align}
where the coefficient matrix $R_{i}$ for the category $i$ is easy to get. Usually, the relations are not independent. To extract the independent ones, we transform the matrix $R_{i}$ to the reduced row echelon form (row canonical form) $S_i$. The rank of $R_{i}$ or $S_{i}$ is equal to the number of independent linear relations and each nonzero row-vector of $S_{i}$ gives a linear relation. That is, the independent constraint equations read
\begin{align}
\sum_{k}S_{i,jk}E_{i,k}=0.
\end{align}
With these constrains, Eq. \eqref{lr0} can be revised to the form
\begin{eqnarray}
D_{i,j}=\sum_{k}A'_{i,jk}E_{i,k},\label{lr1}
\end{eqnarray}
where the matrix $A'_{i}$ is from the matrices $A_{i}$ and $S_{i}$ after all linear dependent constraints are removed.

Third, one extracts the independent terms. Now, the independent terms in $D_{i}$ can be obtained with the help of $A'_{i,jk}$. They correspond to the elements derived with the independent rows of $A'_{i}$ or the independent columns of $A^{\prime T}_{i}$. Similar to the processing of Eq. \eqref{rls}, one transforms the matrix $A^{\prime T}_{i}$ to the reduced row echelon form of $A^{\prime T}_{i}$. Then the labels of the independent terms in $D_i$ and thus the final results can be extracted. The standard form defined in the subsection \ref{redm} ensures that all the linear relations have been used and all the independent monomials of $E_{i,k}$ are really independent.

Fourth, one constructs the contact terms. Because the constraint relations may connect monomials in different categories in this case, we collect all the $D_{i,j}$ and $E_{i,k}$ in two big column vectors $D'_{j}$ and $E'_{k}$, respectively. By repeating the same steps from Eq. \eqref{lr0} to Eq. \eqref{lr1}, one gets the independent terms containing contact terms.

Finally, according to the hermiticity, one has to add an extra $i$ to some terms to ensure that the LECs are real. The Lagrangian with the original building blocks is also recovered with Eq. \eqref{rdb}.

\section{Results and discussions}\label{results}

With the steps given above, we obtain the minimal three-flavor meson-baryon CL to the order ${\mathcal O}(p^4)$. As a cross check, we have confirmed the meson-baryon CLs obtained in Refs. \cite{pin4,pib31,pib32}.

The ${\mathcal O}(p^4)$ meson-baryon CL has the form
\begin{align}
\mathscr{L}^{(4)}_{\mathrm{MB}}=\sum_{n=1}^{540}c_n O_n,
\end{align}
where $c_n$'s are the LECs, and $O_n$'s are the independent chirally invariant terms listed in Table \ref{p4rtab}. The last three terms are contact terms. In Table \ref{pos}, we show processes with the minimal number of mesons and photons to which the ${\mathcal O}(p^4)$ Lagrangian may contribute. The labels of related $O_n$ terms are also given.

\begin{table*}[!h]
\caption{\label{pos}The processes with the minimal number of mesons and photons to which the $\mathcal{O}(p^4)$ monomials may contribute. The numbers in the second column denote the labels of the monomials in Table \ref{p4rtab}.}
\begin{tabular}{r@{$\to$}ll}
\hline\hline
\multicolumn{2}{c}{Process}  &$n$ \\ \hline
$B$ & $B$ & $471\sim473 ,\; 481\sim487 ,\; 538$\\
$B+\gamma$ & $B$ & $406\sim407 ,\; 474\sim480$\\
$B+\gamma$ & $B+\gamma$ & $408\sim419$\\
$B+\gamma$ & $B+3\gamma$ & $539\sim540$\\
$B+M$ & $B$ & $464\sim467 ,\; 469\sim470$\\
$B+M$ & $B+\gamma$ & $364\sim371 ,\; 376\sim393 ,\; 397\sim398 ,\; 401\sim405 ,\; 462\sim463 ,\; 468 ,\; 525\sim531$\\
$B+M$ & $B+2\gamma$ & $354\sim363 ,\; 372\sim375 ,\; 394\sim396 ,\; 399\sim400$\\
$B+M$ & $B+M$ & $218\sim222 ,\; 236\sim243 ,\; 259\sim260 ,\; 266\sim268 ,\; 420\sim461 ,\; 507\sim516 ,\; 518\sim524 ,\; 532\sim537$\\
$B+M$ & $B+M+\gamma$ & $216\sim217 ,\; 228\sim235 ,\; 244\sim255 ,\; 258 ,\; 263\sim265 ,\; 269\sim353 ,\; 505\sim506 ,\; 517$\\
$B+M$ & $B+M+2\gamma$ & $211\sim215 ,\; 223\sim227 ,\; 256\sim257 ,\; 261\sim262$\\
$B+M$ & $B+2M$ & $125\sim148 ,\; 165\sim168 ,\; 173\sim184 ,\; 193\sim196 ,\; 206\sim207 ,\; 209\sim210 ,\; 488\sim504$\\
$B+M$ & $B+2M+\gamma$ & $93\sim124 ,\; 149\sim164 ,\; 169\sim172 ,\; 185\sim192 ,\; 197\sim205 ,\; 208$\\
$B+M$ & $B+3M$ & $1\sim92$\\ \hline\hline
\end{tabular}
\end{table*}

Besides the results in Refs. \cite{pin4,pib31,pib32}, we also check our calculation through other approaches. The independent terms in $D_{i}$ are $C$ and h. c. invariant. Some of them contain two parts as shown in Table \ref{p4rtab}, e.g. $O_4$. The relative phase between them is only $+1$ or $-1$. However, the monomials in $E_{i}$ need not be $C$ or h. c. invariant in the calculation. This property is used to check the correctness of the matrices $A_i$ which must be suitable to keep the $C$ and h. c. invariance on the right hand side of Eq. \eqref{lr0}. It requires that the coefficients of some pairs in $E_{i}$ are equal or only a minus sign difference. In addition, a small mistake in the matrices $R_i$ would also break the $C$ and h. c. invariance of $D_{i,j}$ in Eq. \eqref{lr1}. It will generate confusing results, e,g. giving a very large or very small number of independent terms compared to the lower order Lagrangians or two-flavor $\pi$-nucleon CL of Refs. \cite{pin4,pib31,pib32}.

It seems that 540 is a too large number for independent terms at the order ${\mathcal O}(p^4)$. However, one could not find more relations to reduce this number. Recall that the number of independent normal terms for the $SU(2)$ ($SU(3)$) meson CL at the orders ${\mathcal O}(p^2)$, ${\mathcal O}(p^4)$, and ${\mathcal O}(p^6)$ are 2, 10, and 56 (2, 12, and 94), respectively \cite{,GS1,GS2,p62,p6p}. The increasing number to high orders in the three-flavor case is larger than that in the two-flavor case. In the baryon sector, the numbers of independent terms (the term ($i\, /\!\!\!\! D-m$) not counted) at the orders ${\mathcal O}(p^1)$, ${\mathcal O}(p^2)$, ${\mathcal O}(p^3)$, and ${\mathcal O}(p^4)$ are 1, 7, 23, and 118, respectively, in the $SU(2)$ case. Those in the $SU(3)$ case are 2, 16, 78, and 540, respectively. (Note that the traceless vector and axial vector external sources are adopted in the latter case.) The increasing number in the $SU(3)$ case is much larger. Thus, the number 540 at the fourth chiral order is not so surprising. For a special process, from Table \ref{pos}, only parts of terms and the determination of their coefficients are needed. In reality, the number of independent parameters should be much less than the number of terms shown here. On the other side, the constraint of LECs at this order is possible with further studies or the development of non-perturbative methods, e.g. lattice QCD.

In studying low-energy meson-baryon interactions with chiral perturbation theory, one usually needs to explore the convergence of the chiral expansion. However, the high order correction is not the unique source to improve the expansion. It has been shown that the inclusion of decuplet baryons is also important (see e.g. Ref. \cite{Fettes:2000bb}). The obtained Lagrangian may be used to answer which effect is more important, high order corrections or excited baryon contributions, in specific processes in future investigations.

\section{Summary}\label{summ}
In this paper, we present a systematic and mechanized method for the construction of baryon chiral Lagrangians, which is suitable for computer realization. In the construction, only the independent constraint relations for the chirally-invariant monomials are considered, which remarkably reduces the computational complexity. We have gotten the $SU(3)$ meson-baryon chiral Lagrangian at the order ${\mathcal O}(p^4)$. Now, all the Lorentz-invariant meson-baryon chiral Lagrangians for the one-loop calculation are obtained. Although the number of independent terms is large, only parts of these terms are needed for a special process in which one is interested. We hope that the present work is helpful for further studies on the convergence of the chiral expansion, LEC determinations, model constructions, and so on.

\section*{Acknowledgments}
We thank Professor Qing Wang for helpful discussions. This work was supported by the National Science Foundation of China (NSFC) under Grants No. 11565004, No. 11275115 and the Foundation of Guangxi Key Laboratory for Relativistic Astrophysics.

\appendix
\section{$\gamma$ matrix and EOM}\label{aeom}
This appendix gives a brief introduction to the EOM constraints on the meson-baryon CL. The types of $\gamma$ matrices appearing in the Lagrangian are constrained. One may find detailed descriptions in Refs. \cite{pin3b,pin4}. Although the discussions are for the two-flavor case there, the results are the same as the present case. In this appendix, we also introduce several new relations.

The lowest order EOM from the $SU(3)$ meson-baryon CL is \cite{pib31}
\begin{align}
&i\slashed{D} B-M_0 B+\frac{F}{2}\gamma^\mu\gamma_5[u_\mu,B]+\frac{D}{2}\gamma^\mu\gamma_5\left(\{u_\mu,B\}
-\frac{1}{3}\la\{u_\mu,B\}\ra\right)=0~,\label{pineom}
\end{align}
where $M_0$ is the octet baryon mass in the chiral limit, $F$ and $D$ are the axial-vector coupling constants in the $\mathcal{O}(p^1)$ order CL. The equation means that
\begin{align}
(i\slashed{D}-M_0)B=\mathcal{O}(p^1),
\end{align}
which is similar to the lowest order EOM from the pion-nucleon CL in Ref. \cite{pin4},
\begin{align}
(i\slashed{D}-m)\Psi=\mathcal{O}(p^1).
\end{align}
In other words, we can borrow the relations in the Appendix A of Ref. \cite{pin4} directly, with the replacements $\Psi\to B$ and $m\to M_0$.
For convenience, we collect the constraints on the structure of $\Theta_{\mu\nu\cdots}=\Gamma\times(1\text{ or }\varepsilon^{\alpha\beta\rho\tau} )\times(1\text{ or derivatives }D)$ in Eq. (\ref{clform}) from the baryon EOM as follows:
\begin{enumerate}
\item \label{dd} The case $\Gamma=1$ can give a relation $D^2B=-m^2B+\mathcal{O}(p^1)$, which is similar to the Klein-Gordon equation. From this, one understands that the Lorentz indices of the covariant derivations acting on the baryon field $B$ should be completely different. The existence of the Levi-Civita tensor $\varepsilon^{\alpha\beta\rho\tau}$ is allowed.

\item The case $\Gamma=\gamf$ gives high order terms and it should not exist solely in the Lagrangian.

\item In the case $\Gamma=\gamma^\mu$, one can change it to $iD^\mu$ up to high order terms and it should not appear solely in the Lagrangian, either. The Levi-Civita tensor is allowed, but the structure has been implied in the case $\Gamma=1$.

\item The case $\Gamma=\sigma^{\mu\nu}$ is a little complicated. The contraction of one or two indices of $\sigma^{\mu\nu}$ with those of the covariant derivations acting on the baryon field $B$ gives high order terms or zero. Therefore, $\sigma^{\mu\nu}$ and the derivatives should have completely different Lorentz indices in the allowed monomials. When the Levi-Civita tensor exists, the structure $\sigma_{\mu\nu}\varepsilon^{\mu\nu\lambda\rho}=2i\gamma_5\sigma^{\lambda\rho}$ can be converted to the form of $(\gamma_5\gamma^\lambda D^\rho-\gamma_5\gamma^\rho D^\lambda)$ up to high order terms, which is the following case (v). Because  $\sigma_{\mu\nu}\varepsilon^{\nu\lambda\rho\tau}=(ig_\mu^\lambda\gamma_5\sigma^{\rho\tau}-ig_\mu^\rho\gamma_5\sigma^{\lambda\tau}+ig_\mu^\tau\gamma_5\sigma^{\lambda\rho})$ and the structure $\sigma_{\mu\nu}\varepsilon^{\alpha\beta\rho\tau}$ can also be converted to the form $(g^\alpha_\mu g^\beta_\nu\gamma_5\gamma^\tau D^\rho+\cdots)$ up to high order terms, the independent monomials should not contain any combinations of $\sigma^{\mu\nu}$ and $\varepsilon^{\alpha\beta\rho\tau}$.\label{gammar}

\item In the case $\Gamma=\gamma_5\gamma^\mu$, the Lorentz index should be different from that of any covariant derivative acting on the baryon field $B$. In addition, up to high order terms, the structure $\fgamma_\mu\varepsilon^{\mu\nu\lambda\rho}$ can be converted to the form of $(\sigma^{\nu\lambda}D^\rho+\sigma^{\nu\rho}D^\lambda+\sigma^{\lambda\rho}D^\nu)$ and $\gamma_5\gamma^\mu \epsilon_{\alpha\beta\rho\tau}$ to $(g_\alpha^\mu\sigma_{\beta\rho}D_\tau+\cdots)$, which has been incorporated in the case $\Gamma=\sigma^{\mu\nu}$. Therefore, any combinations of $\fgamma_\mu$ and $\epsilon_{\alpha\beta\rho\tau}$ should not exist in the minimal Lagrangian, either. \label{sigmar}

\end{enumerate}

To summarize, $\Gamma$ can be only $1$, $\fgamma^{\mu}$ or $\sigma^{\mu\nu}$, and their indices should be different from those of covariant derivatives acting on the baryon field $B$. The Levi-Civita tensor $\varepsilon^{\mu\nu\lambda\rho}$ exists only when $\Gamma=1$. To the $\mathcal{O}(p^4)$ order, $\Theta_{\mu\nu\cdots}$ has the forms in Eq. (A.21) of Ref. \cite{pin4} because the number of independent Lorentz indices of $\Theta_{\mu\nu\cdots}$ should be no more than four. Explicitly, they are
\begin{eqnarray}\label{allowedTheta}
&1,\quad D_\mu,\quad D_{\mu\nu},\quad D_{\mu\nu\alpha},\quad D_{\mu\nu\alpha\beta},&\nonumber\\ &\varepsilon_{\mu\nu\alpha\beta}, \quad \varepsilon_{\mu\nu\alpha\tau}D^\tau,\quad \varepsilon_{\mu\nu\alpha\tau}{{D^\tau}}_\beta,&\nonumber\\
&\sigma_{\mu\nu},\quad \sigma_{\mu\nu}D_\alpha,\quad \sigma_{\mu\nu}D_{\alpha\beta},&\nonumber\\
&\gamma_5\gamma_\mu,\quad \gamma_5\gamma_\mu D_\nu,\quad \gamma_5\gamma_\mu D_{\nu\alpha}, \quad \gamma_5\gamma_\mu D_{\nu\alpha\beta}.&
\end{eqnarray}
Other types of structures can be reduced to these forms. A simple method for the reduction is in the heavy baryon formalism. For example, from $\frac{1+/\!\!\! v}{2}\gamma^\alpha\sigma^{\mu\nu}\frac{1+/\!\!\! v}{2}=\frac{1+/\!\!\! v}{2}(v^\alpha\sigma^{\mu\nu}-v^\mu\sigma^{\alpha\nu}+v^\nu\sigma^{\alpha\mu}-iv^\mu g^{\alpha\nu}+iv^\nu g^{\alpha\mu})$, one understands that the structure with $\Theta=\gamma^\alpha\sigma^{\mu\nu}$ may be reduced to the forms like $\sigma^{\mu\nu}D^\alpha$ and $D^\mu$.

The story is not over yet. When constraining the allowed structures of $\Theta_{\mu\nu\cdots}$, we used only the baryon EOM. One may also combine the EOM with the other relations in Subsection \ref{lr} to get new constraints.
From the above items (\ref{gammar}) and (\ref{sigmar}), we have
\begin{align}
\varepsilon^{\cdots}\Bb A^{\cdots}\fgamma^{\cdots} D_{\cdots}\B&\doteq(\Bb A^{\cdots}\sigma^{\cdots} D_{\cdots}\B+\cdots)~,\label{fgamtosigm}\\
\varepsilon^{\cdots}\Bb A^{\cdots}\sigma^{\cdots} D_{\cdots}\B&\doteq(\Bb A^{\cdots}\fgamma^{\cdots} D_{\cdots}\B+\cdots)~,\label{sigmtofgam}
\end{align}
where the Lorentz indices, some constants, and the right-hand-side terms having similar structure are ignored.
By applying Schouten's identity to the left-hand-side terms, we obtain relations for the terms of the $\fgamma$ type and the $\sigma$ type. The independent ones are
\begin{align}
0&\doteq\Bb{\As^{\mu\nu\lambda\rho}}_{\mu\nu}\sigma_{\lambda\rho}{D_{\delta}}^{\delta}\B
+\mathrm{P}(\mu,\nu,\lambda,\rho,\delta),\label{mrl1}\\
0&\doteq\Bb{{\Ase^{\mu\nu\lambda\rho}}_{\mu\nu}}^{\delta}\sigma_{\lambda\rho}D_{\delta}\B
+\mathrm{P}(\mu,\nu,\lambda,\rho,\delta),\\
0&\doteq\Bb{\Ase^{\mu\nu\lambda\rho}}_{\mu\nu\lambda}\gamf \gamma_{\rho}{D_{\delta}}^{\delta}\B
+\mathrm{P}(\mu,\nu,\lambda,\rho,\delta),\\
0&\doteq\Bb{{\Ae^{\mu\nu\lambda\rho}}_{\mu\nu\lambda}}^{\delta}\gamf \gamma_{\rho}D_{\delta}\B
+\mathrm{P}(\mu,\nu,\lambda,\rho,\delta).\label{mrl4}
\end{align}
Here, $\mathrm{P}$ means all permutations of the subscripts behind it. Note that an odd permutation gives a minus sign. These relations were not given in Ref. \cite{pin4}. Fortunately, they only have effects on terms not lower than $\mathcal{O}(p^6)$. By combining the baryon EOM with the partial integration, one obtains three relations similar to Eqs. (A.18)-(A.20) of Ref. \cite{pin4}. In short, we do not obtain any new relations except for Eqs. \eqref{mrl1}-\eqref{mrl4}.

\section{Meson-baryon chiral Lagrangian at order ${\mathcal O}(p^4)$}\label{resp4}
\begin{longtable}{llllll}
\caption{\label{p4rtab}Terms in the $\mathcal{O}(p^4)$ meson-baryon chiral Lagrangian.}\\
\hline\hline $n$ & $O_n$ & $n$ & $O_n$ & $n$ & $O_n$\\
\hline\endfirsthead

\hline\hline $n$ & $O_n$ & $n$ & $O_n$ & $n$ & $O_n$\\
\hline\endhead

\hline\hline 
\endfoot

\hline\endlastfoot

1 & $\la\Bb B u^{\mu}u_{\mu}u^{\nu}u_{\nu}\ra$ &181 & $\la\Bb u^{\mu}h^{\nu\lambda}\gamf \gamma^{\rho}D_{\mu\nu\lambda}B  u_{\rho}\ra+\mathrm{h.c.}$ &361 & $i\la\Bb f_{-}^{\mu\nu}\gamf \gamma^{\lambda}D_{\mu}B  f_{+\nu\lambda}\ra$  \\
2 & $\la\Bb B u^{\mu}u^{\nu}u_{\mu}u_{\nu}\ra$ &182 & $\la\Bb u^{\mu}u^{\nu}h^{\lambda\rho}\gamf \gamma_{\mu}D_{\nu\lambda\rho}B \ra+\mathrm{h.c.}$ &362 & $i\la\Bb f_{+}^{\mu\nu}{f_{-\mu}}^{\lambda}\gamf \gamma_{\nu}D_{\lambda}B \ra+\mathrm{h.c.}$  \\
3 & $\la\Bb B u^{\mu}u^{\nu}u_{\nu}u_{\mu}\ra$ &183 & $\la\Bb u^{\mu}u^{\nu}h^{\lambda\rho}\gamf \gamma_{\nu}D_{\mu\lambda\rho}B \ra+\mathrm{h.c.}$ &363 & $i\la\Bb f_{+}^{\mu\nu}{f_{-\mu}}^{\lambda}\gamf \gamma_{\lambda}D_{\nu}B \ra+\mathrm{h.c.}$  \\
4 & $\la\Bb u^{\mu}B u_{\mu}u^{\nu}u_{\nu}\ra+\mathrm{h.c.}$ &184 & $\la\Bb u^{\mu}u^{\nu}h^{\lambda\rho}\gamf \gamma_{\lambda}D_{\mu\nu\rho}B \ra+\mathrm{h.c.}$ &364 & $i\la\Bb\gamf \gamma^{\mu}D^{\nu}B {f_{+\mu}}^{\lambda}h_{\nu\lambda}\ra+\mathrm{h.c.}$  \\
5 & $\la\Bb u^{\mu}B u^{\nu}u_{\mu}u_{\nu}\ra$ &185 & $\epsilon^{\mu\nu\lambda\rho}\la\Bb B u_{\mu}\ra\la u_{\nu}f_{-\lambda\rho}\ra$ &365 & $i\la\Bb\gamf \gamma^{\mu}D^{\nu}B {f_{+\nu}}^{\lambda}h_{\mu\lambda}\ra+\mathrm{h.c.}$  \\
6 & $\la\Bb u^{\mu}u_{\mu}B u^{\nu}u_{\nu}\ra$ &186 & $\epsilon^{\mu\nu\lambda\rho}\la\Bb u_{\mu}f_{-\nu\lambda}\ra\la u_{\rho}B\ra+\mathrm{h.c.}$ &366 & $i\la\Bb f_{+}^{\mu\nu}\gamf \gamma_{\mu}D^{\lambda}B  h_{\nu\lambda}\ra$  \\
7 & $\la\Bb u^{\mu}u^{\nu}B u_{\mu}u_{\nu}\ra$ &187 & $\epsilon^{\mu\nu\lambda\rho}\la\Bb f_{-\mu\nu}u_{\lambda}\ra\la u_{\rho}B\ra+\mathrm{h.c.}$ &367 & $i\la\Bb f_{+}^{\mu\nu}\gamf \gamma^{\lambda}D_{\mu}B  h_{\nu\lambda}\ra$  \\
8 & $\la\Bb u^{\mu}u^{\nu}B u_{\nu}u_{\mu}\ra$ &188 & $\la\Bb\gamf \gamma^{\mu}D^{\nu}B \ra\la u_{\mu}u^{\lambda}f_{-\nu\lambda}\ra+\mathrm{h.c.}$ &368 & $i\la\Bb h^{\mu\nu}\gamf \gamma_{\mu}D^{\lambda}B  f_{+\nu\lambda}\ra$  \\
9 & $\la\Bb u^{\mu}u_{\mu}u^{\nu}B u_{\nu}\ra+\mathrm{h.c.}$ &189 & $\la\Bb u^{\mu}{f_{-\mu}}^{\nu}\ra\la u^{\lambda}\gamf \gamma_{\lambda}D_{\nu}B \ra+\mathrm{h.c.}$ &369 & $i\la\Bb h^{\mu\nu}\gamf \gamma^{\lambda}D_{\mu}B  f_{+\nu\lambda}\ra$  \\
10 & $\la\Bb u^{\mu}u^{\nu}u_{\mu}B u_{\nu}\ra$ &190 & $\la\Bb\gamf \gamma^{\mu}D^{\nu}B \ra\la u_{\nu}u^{\lambda}f_{-\mu\lambda}\ra+\mathrm{h.c.}$ &370 & $i\la\Bb f_{+}^{\mu\nu}{h_{\mu}}^{\lambda}\gamf \gamma_{\nu}D_{\lambda}B \ra+\mathrm{h.c.}$  \\
11 & $\la\Bb u^{\mu}u_{\mu}u^{\nu}u_{\nu}B\ra$ &191 & $\la\Bb u^{\mu}{f_{-\mu}}^{\nu}\ra\la u^{\lambda}\gamf \gamma_{\nu}D_{\lambda}B \ra+\mathrm{h.c.}$ &371 & $i\la\Bb f_{+}^{\mu\nu}{h_{\mu}}^{\lambda}\gamf \gamma_{\lambda}D_{\nu}B \ra+\mathrm{h.c.}$  \\
12 & $\la\Bb u^{\mu}u^{\nu}u_{\mu}u_{\nu}B\ra$ &192 & $\la\Bb u^{\mu}f_{-}^{\nu\lambda}\ra\la u_{\mu}\gamf \gamma_{\nu}D_{\lambda}B \ra+\mathrm{h.c.}$ &372 & $i\epsilon^{\mu\nu\lambda\rho}\la\Bb{D_{\mu}}^{\sigma}B  f_{+\nu\lambda}f_{-\rho\sigma}\ra+\mathrm{h.c.}$  \\
13 & $\la\Bb u^{\mu}u^{\nu}u_{\nu}u_{\mu}B\ra$ &193 & $\la\Bb\gamf \gamma^{\mu}D^{\nu}B \ra\la u_{\mu}u^{\lambda}h_{\nu\lambda}\ra+\mathrm{h.c.}$ &373 & $i\epsilon^{\mu\nu\lambda\rho}\la\Bb{D_{\mu}}^{\sigma}B  f_{+\nu\sigma}f_{-\lambda\rho}\ra+\mathrm{h.c.}$  \\
14 & $i\la\Bb\sigma^{\mu\nu}B u_{\mu}u_{\nu}u^{\lambda}u_{\lambda}\ra+\mathrm{h.c.}$ &194 & $\la\Bb u^{\mu}{h_{\mu}}^{\nu}\ra\la u^{\lambda}\gamf \gamma_{\lambda}D_{\nu}B \ra+\mathrm{h.c.}$ &374 & $i\epsilon^{\mu\nu\lambda\rho}\la\Bb f_{+\mu\nu}{f_{-\lambda}}^{\sigma}D_{\rho\sigma}B \ra+\mathrm{h.c.}$  \\
15 & $i\la\Bb\sigma^{\mu\nu}B u_{\mu}u^{\lambda}u_{\nu}u_{\lambda}\ra+\mathrm{h.c.}$ &195 & $\la\Bb\gamf \gamma^{\mu}D^{\nu}B \ra\la u_{\nu}u^{\lambda}h_{\mu\lambda}\ra+\mathrm{h.c.}$ &375 & $i\epsilon^{\mu\nu\lambda\rho}\la\Bb{f_{+\mu}}^{\sigma}f_{-\nu\lambda}D_{\rho\sigma}B \ra+\mathrm{h.c.}$  \\
16 & $i\la\Bb\sigma^{\mu\nu}B u_{\mu}u^{\lambda}u_{\lambda}u_{\nu}\ra$ &196 & $\la\Bb u^{\mu}{h_{\mu}}^{\nu}\ra\la u^{\lambda}\gamf \gamma_{\nu}D_{\lambda}B \ra+\mathrm{h.c.}$ &376 & $i\epsilon^{\mu\nu\lambda\rho}\la\Bb{D_{\mu}}^{\sigma}B  f_{+\nu\lambda}h_{\rho\sigma}\ra+\mathrm{h.c.}$  \\
17 & $i\la\Bb\sigma^{\mu\nu}B u^{\lambda}u_{\mu}u_{\nu}u_{\lambda}\ra$ &197 & $\epsilon^{\mu\nu\lambda\rho}\la\Bb{D_{\mu}}^{\sigma}B  u_{\nu}\ra\la u_{\lambda}f_{-\rho\sigma}\ra$ &377 & $i\epsilon^{\mu\nu\lambda\rho}\la\Bb f_{+\mu\nu}{h_{\lambda}}^{\sigma}D_{\rho\sigma}B \ra+\mathrm{h.c.}$  \\
18 & $i\la\Bb u^{\mu}{\sigma_{\mu}}^{\nu}B u_{\nu}u^{\lambda}u_{\lambda}\ra+\mathrm{h.c.}$ &198 & $\epsilon^{\mu\nu\lambda\rho}\la\Bb u_{\mu}{f_{-\nu}}^{\sigma}\ra\la u_{\lambda}D_{\rho\sigma}B \ra+\mathrm{h.c.}$ &378 & $i\la\Bb\gamf \gamma^{\mu}D^{\nu\lambda\rho}B  f_{+\mu\nu}h_{\lambda\rho}\ra+\mathrm{h.c.}$  \\
19 & $i\la\Bb u^{\mu}\sigma^{\nu\lambda}B u_{\mu}u_{\nu}u_{\lambda}\ra+\mathrm{h.c.}$ &199 & $\epsilon^{\mu\nu\lambda\rho}\la\Bb{D_{\mu}}^{\sigma}B \ra\la u_{\nu}u_{\sigma}f_{-\lambda\rho}\ra+\mathrm{h.c.}$ &379 & $i\la\Bb f_{+}^{\mu\nu}\gamf \gamma_{\mu}{D_{\nu}}^{\lambda\rho}B  h_{\lambda\rho}\ra$  \\
20 & $i\la\Bb u^{\mu}\sigma^{\nu\lambda}B u_{\nu}u_{\mu}u_{\lambda}\ra$ &200 & $\epsilon^{\mu\nu\lambda\rho}\la\Bb{D_{\mu}}^{\sigma}B  u_{\nu}\ra\la u_{\sigma}f_{-\lambda\rho}\ra$ &380 & $i\la\Bb h^{\mu\nu}\gamf \gamma^{\lambda}{D_{\mu\nu}}^{\rho}B  f_{+\lambda\rho}\ra$  \\
21 & $i\la\Bb u^{\mu}u_{\mu}\sigma^{\nu\lambda}B u_{\nu}u_{\lambda}\ra$ &201 & $\epsilon^{\mu\nu\lambda\rho}\la\Bb u^{\sigma}f_{-\mu\nu}\ra\la u_{\lambda}D_{\rho\sigma}B \ra+\mathrm{h.c.}$ &381 & $i\la\Bb f_{+}^{\mu\nu}h^{\lambda\rho}\gamf \gamma_{\mu}D_{\nu\lambda\rho}B \ra+\mathrm{h.c.}$  \\
22 & $i\la\Bb u^{\mu}u^{\nu}\sigma_{\mu\nu}B u^{\lambda}u_{\lambda}\ra$ &202 & $\epsilon^{\mu\nu\lambda\rho}\la\Bb f_{-\mu\nu}\ra\la u_{\lambda}u^{\sigma}D_{\rho\sigma}B \ra+\mathrm{h.c.}$ &382 & $i\la\Bb\gamf \gamma^{\mu}D^{\nu}B \nabla_{\mu}{f_{+\nu}}^{\lambda}u_{\lambda}\ra+\mathrm{h.c.}$  \\
23 & $i\la\Bb u^{\mu}u^{\nu}{\sigma_{\mu}}^{\lambda}B u_{\nu}u_{\lambda}\ra+\mathrm{h.c.}$ &203 & $\epsilon^{\mu\nu\lambda\rho}\la\Bb u_{\mu}f_{-\nu\lambda}\ra\la u^{\sigma}D_{\rho\sigma}B \ra+\mathrm{h.c.}$ &383 & $i\la\Bb\gamf \gamma^{\mu}D^{\nu}B \nabla^{\lambda}f_{+\mu\lambda}u_{\nu}\ra+\mathrm{h.c.}$  \\
24 & $i\la\Bb u^{\mu}u^{\nu}{\sigma_{\mu}}^{\lambda}B u_{\lambda}u_{\nu}\ra+\mathrm{h.c.}$ &204 & $\epsilon^{\mu\nu\lambda\rho}\la\Bb f_{-\mu\nu}u^{\sigma}\ra\la u_{\lambda}D_{\rho\sigma}B \ra+\mathrm{h.c.}$ &384 & $i\la\Bb\gamf \gamma^{\mu}D^{\nu}B \nabla^{\lambda}f_{+\nu\lambda}u_{\mu}\ra+\mathrm{h.c.}$  \\
25 & $i\la\Bb u^{\mu}u_{\mu}u^{\nu}{\sigma_{\nu}}^{\lambda}B u_{\lambda}\ra+\mathrm{h.c.}$ &205 & $\epsilon^{\mu\nu\lambda\rho}\la\Bb{f_{-\mu}}^{\sigma}u_{\nu}\ra\la u_{\lambda}D_{\rho\sigma}B \ra+\mathrm{h.c.}$ &385 & $i\la\Bb\nabla^{\mu}{f_{+\mu}}^{\nu}\gamf \gamma_{\nu}D^{\lambda}B  u_{\lambda}\ra$  \\
26 & $i\la\Bb u^{\mu}u^{\nu}u^{\lambda}\sigma_{\mu\nu}B u_{\lambda}\ra+\mathrm{h.c.}$ &206 & $\epsilon^{\mu\nu\lambda\rho}\la\Bb{D_{\mu}}^{\sigma}B  u_{\nu}\ra\la u_{\lambda}h_{\rho\sigma}\ra$ &386 & $i\la\Bb\nabla^{\mu}{f_{+\mu}}^{\nu}\gamf \gamma^{\lambda}D_{\nu}B  u_{\lambda}\ra$  \\
27 & $i\la\Bb u^{\mu}u^{\nu}u^{\lambda}\sigma_{\mu\lambda}B u_{\nu}\ra$ &207 & $\epsilon^{\mu\nu\lambda\rho}\la\Bb u_{\mu}{h_{\nu}}^{\sigma}\ra\la u_{\lambda}D_{\rho\sigma}B \ra+\mathrm{h.c.}$ &387 & $i\la\Bb\nabla^{\mu}f_{+}^{\nu\lambda}\gamf \gamma_{\mu}D_{\nu}B  u_{\lambda}\ra$  \\
28 & $i\la\Bb u^{\mu}u_{\mu}u^{\nu}u^{\lambda}\sigma_{\nu\lambda}B\ra+\mathrm{h.c.}$ &208 & $\la\Bb u^{\mu}f_{-}^{\nu\lambda}\ra\la u^{\rho}\gamf \gamma_{\nu}D_{\mu\lambda\rho}B \ra+\mathrm{h.c.}$ &388 & $i\la\Bb u^{\mu}\gamf \gamma_{\mu}D^{\nu}B \nabla^{\lambda}f_{+\nu\lambda}\ra$  \\
29 & $i\la\Bb u^{\mu}u^{\nu}u_{\mu}u^{\lambda}\sigma_{\nu\lambda}B\ra+\mathrm{h.c.}$ &209 & $\la\Bb\gamf \gamma^{\mu}D^{\nu\lambda\rho}B \ra\la u_{\mu}u_{\nu}h_{\lambda\rho}\ra+\mathrm{h.c.}$ &389 & $i\la\Bb u^{\mu}\gamf \gamma^{\nu}D_{\mu}B \nabla^{\lambda}f_{+\nu\lambda}\ra$  \\
30 & $i\la\Bb u^{\mu}u^{\nu}u_{\nu}u^{\lambda}\sigma_{\mu\lambda}B\ra$ &210 & $\la\Bb u^{\mu}h^{\nu\lambda}\ra\la u^{\rho}\gamf \gamma_{\rho}D_{\mu\nu\lambda}B \ra+\mathrm{h.c.}$ &390 & $i\la\Bb u^{\mu}\gamf \gamma^{\nu}D^{\lambda}B \nabla_{\mu}f_{+\nu\lambda}\ra$  \\
31 & $i\la\Bb u^{\mu}u^{\nu}u^{\lambda}u_{\mu}\sigma_{\nu\lambda}B\ra$ &211 & $\la\Bb B f_{-}^{\mu\nu}f_{-\mu\nu}\ra$ &391 & $i\la\Bb\nabla^{\mu}{f_{+\mu}}^{\nu}u^{\lambda}\gamf \gamma_{\nu}D_{\lambda}B \ra+\mathrm{h.c.}$  \\
32 & $\la\Bb D^{\mu\nu}B  u_{\mu}u_{\nu}u^{\lambda}u_{\lambda}\ra+\mathrm{h.c.}$ &212 & $\la\Bb f_{-}^{\mu\nu}B f_{-\mu\nu}\ra$ &392 & $i\la\Bb\nabla^{\mu}{f_{+\mu}}^{\nu}u^{\lambda}\gamf \gamma_{\lambda}D_{\nu}B \ra+\mathrm{h.c.}$  \\
33 & $\la\Bb D^{\mu\nu}B  u_{\mu}u^{\lambda}u_{\nu}u_{\lambda}\ra+\mathrm{h.c.}$ &213 & $\la\Bb f_{-}^{\mu\nu}f_{-\mu\nu}B\ra$ &393 & $i\la\Bb\nabla^{\mu}f_{+}^{\nu\lambda}u_{\mu}\gamf \gamma_{\nu}D_{\lambda}B \ra+\mathrm{h.c.}$  \\
34 & $\la\Bb D^{\mu\nu}B  u_{\mu}u^{\lambda}u_{\lambda}u_{\nu}\ra$ &214 & $i\la\Bb\sigma^{\mu\nu}B{f_{-\mu}}^{\lambda}f_{-\nu\lambda}\ra$ &394 & $i\epsilon^{\mu\nu\lambda\rho}\la\Bb f_{-\mu\nu}\ra\la f_{+\lambda\rho}B\ra+\mathrm{h.c.}$  \\
35 & $\la\Bb D^{\mu\nu}B  u^{\lambda}u_{\mu}u_{\nu}u_{\lambda}\ra$ &215 & $i\la\Bb f_{-}^{\mu\nu}{f_{-\mu}}^{\lambda}\sigma_{\nu\lambda}B\ra$ &395 & $i\la\Bb\gamf \gamma^{\mu}D^{\nu}B \ra\la{f_{+\mu}}^{\lambda}f_{-\nu\lambda}\ra$  \\
36 & $\la\Bb u^{\mu}{D_{\mu}}^{\nu}B  u_{\nu}u^{\lambda}u_{\lambda}\ra+\mathrm{h.c.}$ &216 & $i\la\Bb\sigma^{\mu\nu}B{f_{-\mu}}^{\lambda}h_{\nu\lambda}\ra+\mathrm{h.c.}$ &396 & $i\la\Bb\gamf \gamma^{\mu}D^{\nu}B \ra\la{f_{+\nu}}^{\lambda}f_{-\mu\lambda}\ra$  \\
37 & $\la\Bb u^{\mu}{D_{\mu}}^{\nu}B  u^{\lambda}u_{\nu}u_{\lambda}\ra$ &217 & $i\la\Bb f_{-}^{\mu\nu}{h_{\mu}}^{\lambda}\sigma_{\nu\lambda}B\ra+\mathrm{h.c.}$ &397 & $i\la\Bb\gamf \gamma^{\mu}D^{\nu}B \ra\la{f_{+\mu}}^{\lambda}h_{\nu\lambda}\ra$  \\
38 & $\la\Bb u^{\mu}D^{\nu\lambda}B  u_{\mu}u_{\nu}u_{\lambda}\ra+\mathrm{h.c.}$ &218 & $\la\Bb B h^{\mu\nu}h_{\mu\nu}\ra$ &398 & $i\la\Bb\gamf \gamma^{\mu}D^{\nu}B \ra\la{f_{+\nu}}^{\lambda}h_{\mu\lambda}\ra$  \\
39 & $\la\Bb u^{\mu}D^{\nu\lambda}B  u_{\nu}u_{\mu}u_{\lambda}\ra$ &219 & $\la\Bb h^{\mu\nu}B h_{\mu\nu}\ra$ &399 & $i\epsilon^{\mu\nu\lambda\rho}\la\Bb{f_{-\mu}}^{\sigma}\ra\la f_{+\nu\lambda}D_{\rho\sigma}B \ra+\mathrm{h.c.}$  \\
40 & $\la\Bb u^{\mu}u_{\mu}D^{\nu\lambda}B  u_{\nu}u_{\lambda}\ra$ &220 & $\la\Bb h^{\mu\nu}h_{\mu\nu}B\ra$ &400 & $i\epsilon^{\mu\nu\lambda\rho}\la\Bb f_{-\mu\nu}\ra\la{f_{+\lambda}}^{\sigma}D_{\rho\sigma}B \ra+\mathrm{h.c.}$  \\
41 & $\la\Bb u^{\mu}u^{\nu}D_{\mu\nu}B  u^{\lambda}u_{\lambda}\ra$ &221 & $i\la\Bb\sigma^{\mu\nu}B{h_{\mu}}^{\lambda}h_{\nu\lambda}\ra$ &401 & $i\epsilon^{\mu\nu\lambda\rho}\la\Bb{h_{\mu}}^{\sigma}\ra\la f_{+\nu\lambda}D_{\rho\sigma}B \ra+\mathrm{h.c.}$  \\
42 & $\la\Bb u^{\mu}u^{\nu}{D_{\mu}}^{\lambda}B  u_{\nu}u_{\lambda}\ra+\mathrm{h.c.}$ &222 & $i\la\Bb h^{\mu\nu}{h_{\mu}}^{\lambda}\sigma_{\nu\lambda}B\ra$ &402 & $i\la\Bb\gamf \gamma^{\mu}D^{\nu\lambda\rho}B \ra\la f_{+\mu\nu}h_{\lambda\rho}\ra$  \\
43 & $\la\Bb u^{\mu}u^{\nu}{D_{\mu}}^{\lambda}B  u_{\lambda}u_{\nu}\ra+\mathrm{h.c.}$ &223 & $\la\Bb D^{\mu\nu}B {f_{-\mu}}^{\lambda}f_{-\nu\lambda}\ra$ &403 & $i\la\Bb\gamf \gamma^{\mu}D^{\nu}B \ra\la\nabla_{\mu}{f_{+\nu}}^{\lambda}u_{\lambda}\ra$  \\
44 & $\la\Bb u^{\mu}u_{\mu}u^{\nu}{D_{\nu}}^{\lambda}B  u_{\lambda}\ra+\mathrm{h.c.}$ &224 & $\la\Bb f_{-}^{\mu\nu}{D_{\mu}}^{\lambda}B  f_{-\nu\lambda}\ra$ &404 & $i\la\Bb\gamf \gamma^{\mu}D^{\nu}B \ra\la\nabla^{\lambda}f_{+\mu\lambda}u_{\nu}\ra$  \\
45 & $\la\Bb u^{\mu}u^{\nu}u_{\mu}{D_{\nu}}^{\lambda}B  u_{\lambda}\ra$ &225 & $\la\Bb f_{-}^{\mu\nu}{f_{-\mu}}^{\lambda}D_{\nu\lambda}B \ra$ &405 & $i\la\Bb\gamf \gamma^{\mu}D^{\nu}B \ra\la\nabla^{\lambda}f_{+\nu\lambda}u_{\mu}\ra$  \\
46 & $\la\Bb u^{\mu}u^{\nu}u^{\lambda}D_{\mu\nu}B  u_{\lambda}\ra+\mathrm{h.c.}$ &226 & $i\la\Bb\sigma^{\mu\nu}D^{\lambda\rho}B  f_{-\mu\lambda}f_{-\nu\rho}\ra$ &406 & $\la\Bb\sigma^{\mu\nu}B\nabla_{\mu}\nabla^{\lambda}f_{+\nu\lambda}\ra$  \\
47 & $\la\Bb u^{\mu}u^{\nu}u^{\lambda}D_{\mu\lambda}B  u_{\nu}\ra$ &227 & $i\la\Bb f_{-}^{\mu\nu}f_{-}^{\lambda\rho}\sigma_{\mu\lambda}D_{\nu\rho}B \ra$ &407 & $\la\Bb\nabla^{\mu}\nabla_{\mu}f_{+}^{\nu\lambda}\sigma_{\nu\lambda}B\ra$  \\
48 & $\la\Bb u^{\mu}u_{\mu}u^{\nu}u^{\lambda}D_{\nu\lambda}B \ra+\mathrm{h.c.}$ &228 & $\la\Bb D^{\mu\nu}B {f_{-\mu}}^{\lambda}h_{\nu\lambda}\ra+\mathrm{h.c.}$ &408 & $\la\Bb B f_{+}^{\mu\nu}f_{+\mu\nu}\ra$  \\
49 & $\la\Bb u^{\mu}u^{\nu}u_{\mu}u^{\lambda}D_{\nu\lambda}B \ra+\mathrm{h.c.}$ &229 & $\la\Bb f_{-}^{\mu\nu}{D_{\mu}}^{\lambda}B  h_{\nu\lambda}\ra$ &409 & $\la\Bb f_{+}^{\mu\nu}B f_{+\mu\nu}\ra$  \\
50 & $\la\Bb u^{\mu}u^{\nu}u_{\nu}u^{\lambda}D_{\mu\lambda}B \ra$ &230 & $\la\Bb h^{\mu\nu}{D_{\mu}}^{\lambda}B  f_{-\nu\lambda}\ra$ &410 & $\la\Bb f_{+}^{\mu\nu}f_{+\mu\nu}B\ra$  \\
51 & $\la\Bb u^{\mu}u^{\nu}u^{\lambda}u_{\mu}D_{\nu\lambda}B \ra$ &231 & $\la\Bb f_{-}^{\mu\nu}{h_{\mu}}^{\lambda}D_{\nu\lambda}B \ra+\mathrm{h.c.}$ &411 & $i\la\Bb\sigma^{\mu\nu}B{f_{+\mu}}^{\lambda}f_{+\nu\lambda}\ra$  \\
52 & $i\la\Bb\sigma^{\mu\nu}D^{\lambda\rho}B  u_{\mu}u_{\nu}u_{\lambda}u_{\rho}\ra+\mathrm{h.c.}$ &232 & $i\la\Bb\sigma^{\mu\nu}D^{\lambda\rho}B  f_{-\mu\nu}h_{\lambda\rho}\ra+\mathrm{h.c.}$ &412 & $i\la\Bb f_{+}^{\mu\nu}{f_{+\mu}}^{\lambda}\sigma_{\nu\lambda}B\ra$  \\
53 & $i\la\Bb\sigma^{\mu\nu}D^{\lambda\rho}B  u_{\mu}u_{\lambda}u_{\nu}u_{\rho}\ra+\mathrm{h.c.}$ &233 & $i\la\Bb\sigma^{\mu\nu}D^{\lambda\rho}B  f_{-\mu\lambda}h_{\nu\rho}\ra+\mathrm{h.c.}$ &413 & $\la\Bb D^{\mu\nu}B {f_{+\mu}}^{\lambda}f_{+\nu\lambda}\ra$  \\
54 & $i\la\Bb\sigma^{\mu\nu}D^{\lambda\rho}B  u_{\mu}u_{\lambda}u_{\rho}u_{\nu}\ra$ &234 & $i\la\Bb f_{-}^{\mu\nu}h^{\lambda\rho}\sigma_{\mu\nu}D_{\lambda\rho}B \ra+\mathrm{h.c.}$ &414 & $\la\Bb f_{+}^{\mu\nu}{D_{\mu}}^{\lambda}B  f_{+\nu\lambda}\ra$  \\
55 & $i\la\Bb\sigma^{\mu\nu}D^{\lambda\rho}B  u_{\lambda}u_{\mu}u_{\nu}u_{\rho}\ra$ &235 & $i\la\Bb f_{-}^{\mu\nu}h^{\lambda\rho}\sigma_{\mu\lambda}D_{\nu\rho}B \ra+\mathrm{h.c.}$ &415 & $\la\Bb f_{+}^{\mu\nu}{f_{+\mu}}^{\lambda}D_{\nu\lambda}B \ra$  \\
56 & $i\la\Bb u^{\mu}{\sigma_{\mu}}^{\nu}D^{\lambda\rho}B  u_{\nu}u_{\lambda}u_{\rho}\ra+\mathrm{h.c.}$ &236 & $\la\Bb D^{\mu\nu}B {h_{\mu}}^{\lambda}h_{\nu\lambda}\ra$ &416 & $i\la\Bb\sigma^{\mu\nu}D^{\lambda\rho}B  f_{+\mu\lambda}f_{+\nu\rho}\ra$  \\
57 & $i\la\Bb u^{\mu}\sigma^{\nu\lambda}{D_{\mu}}^{\rho}B  u_{\nu}u_{\lambda}u_{\rho}\ra+\mathrm{h.c.}$ &237 & $\la\Bb h^{\mu\nu}{D_{\mu}}^{\lambda}B  h_{\nu\lambda}\ra$ &417 & $i\la\Bb f_{+}^{\mu\nu}f_{+}^{\lambda\rho}\sigma_{\mu\lambda}D_{\nu\rho}B \ra$  \\
58 & $i\la\Bb u^{\mu}\sigma^{\nu\lambda}{D_{\mu}}^{\rho}B  u_{\nu}u_{\rho}u_{\lambda}\ra$ &238 & $\la\Bb h^{\mu\nu}{h_{\mu}}^{\lambda}D_{\nu\lambda}B \ra$ &418 & $i\la\Bb f_{+}^{\mu\nu}\ra\la{f_{+\mu}}^{\lambda}\sigma_{\nu\lambda}B\ra$  \\
59 & $i\la\Bb u^{\mu}u^{\nu}\sigma_{\mu\nu}D^{\lambda\rho}B  u_{\lambda}u_{\rho}\ra$ &239 & $i\la\Bb\sigma^{\mu\nu}D^{\lambda\rho}B  h_{\mu\lambda}h_{\nu\rho}\ra$ &419 & $i\la\Bb f_{+}^{\mu\nu}\ra\la f_{+}^{\lambda\rho}\sigma_{\mu\lambda}D_{\nu\rho}B \ra$  \\
60 & $i\la\Bb u^{\mu}u^{\nu}{\sigma_{\mu}}^{\lambda}{D_{\nu}}^{\rho}B  u_{\lambda}u_{\rho}\ra+\mathrm{h.c.}$ &240 & $i\la\Bb h^{\mu\nu}h^{\lambda\rho}\sigma_{\mu\lambda}D_{\nu\rho}B \ra$ &420 & $\la\Bb B u^{\mu}u_{\mu}\chip\ra+\mathrm{h.c.}$  \\
61 & $i\la\Bb u^{\mu}u^{\nu}{\sigma_{\mu}}^{\lambda}{D_{\nu}}^{\rho}B  u_{\rho}u_{\lambda}\ra+\mathrm{h.c.}$ &241 & $\la\Bb D^{\mu\nu\lambda\rho}B  h_{\mu\nu}h_{\lambda\rho}\ra$ &421 & $\la\Bb B u^{\mu}\chip u_{\mu}\ra$  \\
62 & $i\la\Bb u^{\mu}u^{\nu}\sigma^{\lambda\rho}D_{\mu\nu}B  u_{\lambda}u_{\rho}\ra$ &242 & $\la\Bb h^{\mu\nu}{D_{\mu\nu}}^{\lambda\rho}B  h_{\lambda\rho}\ra$ &422 & $\la\Bb u^{\mu}B u_{\mu}\chip\ra+\mathrm{h.c.}$  \\
63 & $i\la\Bb u^{\mu}u^{\nu}u^{\lambda}\sigma_{\mu\nu}{D_{\lambda}}^{\rho}B  u_{\rho}\ra+\mathrm{h.c.}$ &243 & $\la\Bb h^{\mu\nu}h^{\lambda\rho}D_{\mu\nu\lambda\rho}B \ra$ &423 & $\la\Bb\chip B u^{\mu}u_{\mu}\ra$  \\
64 & $i\la\Bb u^{\mu}u^{\nu}u^{\lambda}\sigma_{\mu\lambda}{D_{\nu}}^{\rho}B  u_{\rho}\ra$ &244 & $\la\Bb B u^{\mu}\nabla^{\nu}f_{-\mu\nu}\ra+\mathrm{h.c.}$ &424 & $\la\Bb u^{\mu}u_{\mu}B\chip\ra$  \\
65 & $i\la\Bb u^{\mu}u^{\nu}u^{\lambda}{\sigma_{\mu}}^{\rho}D_{\nu\lambda}B  u_{\rho}\ra+\mathrm{h.c.}$ &245 & $\la\Bb u^{\mu}B\nabla^{\nu}f_{-\mu\nu}\ra$ &425 & $\la\Bb u^{\mu}\chip B u_{\mu}\ra+\mathrm{h.c.}$  \\
66 & $i\la\Bb u^{\mu}u^{\nu}u^{\lambda}u^{\rho}\sigma_{\mu\nu}D_{\lambda\rho}B \ra+\mathrm{h.c.}$ &246 & $\la\Bb\nabla^{\mu}{f_{-\mu}}^{\nu}B u_{\nu}\ra$ &426 & $\la\Bb u^{\mu}u_{\mu}\chip B\ra+\mathrm{h.c.}$  \\
67 & $i\la\Bb u^{\mu}u^{\nu}u^{\lambda}u^{\rho}\sigma_{\mu\lambda}D_{\nu\rho}B \ra+\mathrm{h.c.}$ &247 & $\la\Bb u^{\mu}\nabla^{\nu}f_{-\mu\nu}B\ra+\mathrm{h.c.}$ &427 & $\la\Bb u^{\mu}\chip u_{\mu}B\ra$  \\
68 & $i\la\Bb u^{\mu}u^{\nu}u^{\lambda}u^{\rho}\sigma_{\mu\rho}D_{\nu\lambda}B \ra$ &248 & $i\la\Bb\sigma^{\mu\nu}B u_{\mu}\nabla^{\lambda}f_{-\nu\lambda}\ra+\mathrm{h.c.}$ &428 & $i\la\Bb\sigma^{\mu\nu}B u_{\mu}u_{\nu}\chip\ra+\mathrm{h.c.}$  \\
69 & $i\la\Bb u^{\mu}u^{\nu}u^{\lambda}u^{\rho}\sigma_{\nu\lambda}D_{\mu\rho}B \ra$ &249 & $i\la\Bb\sigma^{\mu\nu}B u^{\lambda}\nabla_{\mu}f_{-\nu\lambda}\ra+\mathrm{h.c.}$ &429 & $i\la\Bb\sigma^{\mu\nu}B u_{\mu}\chip u_{\nu}\ra$  \\
70 & $\la\Bb D^{\mu\nu\lambda\rho}B  u_{\mu}u_{\nu}u_{\lambda}u_{\rho}\ra$ &250 & $i\la\Bb u^{\mu}\nabla_{\mu}f_{-}^{\nu\lambda}\sigma_{\nu\lambda}B\ra+\mathrm{h.c.}$ &430 & $i\la\Bb u^{\mu}{\sigma_{\mu}}^{\nu}B u_{\nu}\chip\ra+\mathrm{h.c.}$  \\
71 & $\la\Bb u^{\mu}{D_{\mu}}^{\nu\lambda\rho}B  u_{\nu}u_{\lambda}u_{\rho}\ra$ &251 & $i\la\Bb u^{\mu}\nabla^{\nu}{f_{-\nu}}^{\lambda}\sigma_{\mu\lambda}B\ra+\mathrm{h.c.}$ &431 & $i\la\Bb\chip\sigma^{\mu\nu}B u_{\mu}u_{\nu}\ra$  \\
72 & $\la\Bb u^{\mu}u^{\nu}{D_{\mu\nu}}^{\lambda\rho}B  u_{\lambda}u_{\rho}\ra$ &252 & $\la\Bb D^{\mu\nu}B  u_{\mu}\nabla^{\lambda}f_{-\nu\lambda}\ra+\mathrm{h.c.}$ &432 & $i\la\Bb u^{\mu}u^{\nu}\sigma_{\mu\nu}B\chip\ra$  \\
73 & $\la\Bb u^{\mu}u^{\nu}u^{\lambda}{D_{\mu\nu\lambda}}^{\rho}B  u_{\rho}\ra$ &253 & $\la\Bb u^{\mu}{D_{\mu}}^{\nu}B \nabla^{\lambda}f_{-\nu\lambda}\ra$ &433 & $i\la\Bb u^{\mu}\chip{\sigma_{\mu}}^{\nu}B u_{\nu}\ra+\mathrm{h.c.}$  \\
74 & $\la\Bb u^{\mu}u^{\nu}u^{\lambda}u^{\rho}D_{\mu\nu\lambda\rho}B \ra$ &254 & $\la\Bb\nabla^{\mu}{f_{-\mu}}^{\nu}{D_{\nu}}^{\lambda}B  u_{\lambda}\ra$ &434 & $i\la\Bb u^{\mu}u^{\nu}\chip\sigma_{\mu\nu}B\ra+\mathrm{h.c.}$  \\
75 & $\la\Bb B\ra\la u^{\mu}u_{\mu}u^{\nu}u_{\nu}\ra$ &255 & $\la\Bb u^{\mu}\nabla^{\nu}{f_{-\nu}}^{\lambda}D_{\mu\lambda}B \ra+\mathrm{h.c.}$ &435 & $i\la\Bb u^{\mu}\chip u^{\nu}\sigma_{\mu\nu}B\ra$  \\
76 & $\la\Bb B u^{\mu}\ra\la u_{\mu}u^{\nu}u_{\nu}\ra$ &256 & $\la\Bb B\ra\la f_{-}^{\mu\nu}f_{-\mu\nu}\ra$ &436 & $\la\Bb D^{\mu\nu}B  u_{\mu}u_{\nu}\chip\ra+\mathrm{h.c.}$  \\
77 & $\la\Bb B u^{\mu}u_{\mu}\ra\la u^{\nu}u_{\nu}\ra$ &257 & $i\la\Bb f_{-}^{\mu\nu}\ra\la{f_{-\mu}}^{\lambda}\sigma_{\nu\lambda}B\ra$ &437 & $\la\Bb D^{\mu\nu}B  u_{\mu}\chip u_{\nu}\ra$  \\
78 & $\la\Bb B\ra\la u^{\mu}u^{\nu}u_{\mu}u_{\nu}\ra$ &258 & $i\la\Bb h^{\mu\nu}\ra\la{f_{-\mu}}^{\lambda}\sigma_{\nu\lambda}B\ra+\mathrm{h.c.}$ &438 & $\la\Bb u^{\mu}{D_{\mu}}^{\nu}B  u_{\nu}\chip\ra+\mathrm{h.c.}$  \\
79 & $i\la\Bb\sigma^{\mu\nu}B\ra\la u_{\mu}u_{\nu}u^{\lambda}u_{\lambda}\ra$ &259 & $\la\Bb B\ra\la h^{\mu\nu}h_{\mu\nu}\ra$ &439 & $\la\Bb\chip D^{\mu\nu}B  u_{\mu}u_{\nu}\ra$  \\
80 & $i\la\Bb\sigma^{\mu\nu}B u_{\mu}u_{\nu}\ra\la u^{\lambda}u_{\lambda}\ra$ &260 & $i\la\Bb h^{\mu\nu}\ra\la{h_{\mu}}^{\lambda}\sigma_{\nu\lambda}B\ra$ &440 & $\la\Bb u^{\mu}u^{\nu}D_{\mu\nu}B \chip\ra$  \\
81 & $i\la\Bb u^{\mu}u^{\nu}u_{\nu}\ra\la u^{\lambda}\sigma_{\mu\lambda}B\ra+\mathrm{h.c.}$ &261 & $\la\Bb D^{\mu\nu}B \ra\la{f_{-\mu}}^{\lambda}f_{-\nu\lambda}\ra$ &441 & $\la\Bb u^{\mu}\chip{D_{\mu}}^{\nu}B  u_{\nu}\ra+\mathrm{h.c.}$  \\
82 & $\la\Bb D^{\mu\nu}B \ra\la u_{\mu}u_{\nu}u^{\lambda}u_{\lambda}\ra$ &262 & $i\la\Bb f_{-}^{\mu\nu}\ra\la f_{-}^{\lambda\rho}\sigma_{\mu\lambda}D_{\nu\rho}B \ra$ &442 & $\la\Bb u^{\mu}u^{\nu}\chip D_{\mu\nu}B \ra+\mathrm{h.c.}$  \\
83 & $\la\Bb D^{\mu\nu}B  u_{\mu}\ra\la u_{\nu}u^{\lambda}u_{\lambda}\ra$ &263 & $\la\Bb D^{\mu\nu}B \ra\la{f_{-\mu}}^{\lambda}h_{\nu\lambda}\ra$ &443 & $\la\Bb u^{\mu}\chip u^{\nu}D_{\mu\nu}B \ra$  \\
84 & $\la\Bb D^{\mu\nu}B  u_{\mu}u_{\nu}\ra\la u^{\lambda}u_{\lambda}\ra$ &264 & $i\la\Bb h^{\mu\nu}\ra\la f_{-}^{\lambda\rho}\sigma_{\lambda\rho}D_{\mu\nu}B \ra+\mathrm{h.c.}$ &444 & $\la\Bb B\ra\la u^{\mu}u_{\mu}\chip\ra$  \\
85 & $\la\Bb u^{\mu}u_{\mu}\ra\la u^{\nu}u^{\lambda}D_{\nu\lambda}B \ra+\mathrm{h.c.}$ &265 & $i\la\Bb h^{\mu\nu}\ra\la f_{-}^{\lambda\rho}\sigma_{\mu\lambda}D_{\nu\rho}B \ra+\mathrm{h.c.}$ &445 & $\la\Bb B u^{\mu}\ra\la u_{\mu}\chip\ra$  \\
86 & $\la\Bb D^{\mu\nu}B \ra\la u_{\mu}u^{\lambda}u_{\nu}u_{\lambda}\ra$ &266 & $\la\Bb D^{\mu\nu}B \ra\la{h_{\mu}}^{\lambda}h_{\nu\lambda}\ra$ &446 & $\la\Bb B u^{\mu}u_{\mu}\ra\la\chip\ra$  \\
87 & $\la\Bb u^{\mu}u^{\nu}u_{\mu}\ra\la u^{\lambda}D_{\nu\lambda}B \ra+\mathrm{h.c.}$ &267 & $i\la\Bb h^{\mu\nu}\ra\la h^{\lambda\rho}\sigma_{\mu\lambda}D_{\nu\rho}B \ra$ &447 & $\la\Bb u^{\mu}\chip\ra\la u_{\mu}B\ra+\mathrm{h.c.}$  \\
88 & $i\la\Bb\sigma^{\mu\nu}D^{\lambda\rho}B \ra\la u_{\mu}u_{\nu}u_{\lambda}u_{\rho}\ra$ &268 & $\la\Bb D^{\mu\nu\lambda\rho}B \ra\la h_{\mu\nu}h_{\lambda\rho}\ra$ &448 & $\la\Bb\chip\ra\la u^{\mu}u_{\mu}B\ra+\mathrm{h.c.}$  \\
89 & $i\la\Bb\sigma^{\mu\nu}D^{\lambda\rho}B  u_{\mu}u_{\nu}\ra\la u_{\lambda}u_{\rho}\ra$ &269 & $\la\Bb B\ra\la u^{\mu}\nabla^{\nu}f_{-\mu\nu}\ra$ &449 & $\la\Bb\chip u^{\mu}\ra\la u_{\mu}B\ra+\mathrm{h.c.}$  \\
90 & $i\la\Bb u^{\mu}u^{\nu}u^{\lambda}\ra\la u^{\rho}\sigma_{\mu\rho}D_{\nu\lambda}B \ra+\mathrm{h.c.}$ &270 & $i\la\Bb\nabla^{\mu}{f_{-\mu}}^{\nu}\ra\la u^{\lambda}\sigma_{\nu\lambda}B\ra+\mathrm{h.c.}$ &450 & $\la\Bb u^{\mu}B u_{\mu}\ra\la\chip\ra$  \\
91 & $\la\Bb D^{\mu\nu\lambda\rho}B \ra\la u_{\mu}u_{\nu}u_{\lambda}u_{\rho}\ra$ &271 & $i\la\Bb\nabla^{\mu}f_{-}^{\nu\lambda}\ra\la u_{\nu}\sigma_{\mu\lambda}B\ra+\mathrm{h.c.}$ &451 & $i\la\Bb\sigma^{\mu\nu}B\ra\la u_{\mu}u_{\nu}\chip\ra$  \\
92 & $\la\Bb D^{\mu\nu\lambda\rho}B  u_{\mu}\ra\la u_{\nu}u_{\lambda}u_{\rho}\ra$ &272 & $\la\Bb D^{\mu\nu}B \ra\la u_{\mu}\nabla^{\lambda}f_{-\nu\lambda}\ra$ &452 & $i\la\Bb\sigma^{\mu\nu}B u_{\mu}u_{\nu}\ra\la\chip\ra$  \\
93 & $\epsilon^{\mu\nu\lambda\rho}\la\Bb B u_{\mu}u_{\nu}f_{-\lambda\rho}\ra+\mathrm{h.c.}$ &273 & $i\la\Bb B f_{+}^{\mu\nu}u_{\mu}u_{\nu}\ra+\mathrm{h.c.}$ &453 & $i\la\Bb u^{\mu}\chip\ra\la u^{\nu}\sigma_{\mu\nu}B\ra+\mathrm{h.c.}$  \\
94 & $\epsilon^{\mu\nu\lambda\rho}\la\Bb u_{\mu}B u_{\nu}f_{-\lambda\rho}\ra+\mathrm{h.c.}$ &274 & $i\la\Bb B u^{\mu}{f_{+\mu}}^{\nu}u_{\nu}\ra$ &454 & $i\la\Bb\chip\ra\la u^{\mu}u^{\nu}\sigma_{\mu\nu}B\ra+\mathrm{h.c.}$  \\
95 & $\epsilon^{\mu\nu\lambda\rho}\la\Bb u_{\mu}f_{-\nu\lambda}B u_{\rho}\ra+\mathrm{h.c.}$ &275 & $i\la\Bb f_{+}^{\mu\nu}B u_{\mu}u_{\nu}\ra$ &455 & $\la\Bb D^{\mu\nu}B \ra\la u_{\mu}u_{\nu}\chip\ra$  \\
96 & $\epsilon^{\mu\nu\lambda\rho}\la\Bb u_{\mu}u_{\nu}f_{-\lambda\rho}B\ra+\mathrm{h.c.}$ &276 & $i\la\Bb u^{\mu}B{f_{+\mu}}^{\nu}u_{\nu}\ra+\mathrm{h.c.}$ &456 & $\la\Bb D^{\mu\nu}B  u_{\mu}\ra\la u_{\nu}\chip\ra$  \\
97 & $\la\Bb\gamf \gamma^{\mu}D^{\nu}B  u_{\mu}u^{\lambda}f_{-\nu\lambda}\ra+\mathrm{h.c.}$ &277 & $i\la\Bb f_{+}^{\mu\nu}u_{\mu}B u_{\nu}\ra+\mathrm{h.c.}$ &457 & $\la\Bb D^{\mu\nu}B  u_{\mu}u_{\nu}\ra\la\chip\ra$  \\
98 & $\la\Bb\gamf \gamma^{\mu}D^{\nu}B  u_{\nu}u^{\lambda}f_{-\mu\lambda}\ra+\mathrm{h.c.}$ &278 & $i\la\Bb u^{\mu}u^{\nu}B f_{+\mu\nu}\ra$ &458 & $\la\Bb u^{\mu}\chip\ra\la u^{\nu}D_{\mu\nu}B \ra+\mathrm{h.c.}$  \\
99 & $\la\Bb\gamf \gamma^{\mu}D^{\nu}B  u^{\lambda}u_{\mu}f_{-\nu\lambda}\ra+\mathrm{h.c.}$ &279 & $i\la\Bb f_{+}^{\mu\nu}u_{\mu}u_{\nu}B\ra+\mathrm{h.c.}$ &459 & $\la\Bb\chip\ra\la u^{\mu}u^{\nu}D_{\mu\nu}B \ra+\mathrm{h.c.}$  \\
100 & $\la\Bb\gamf \gamma^{\mu}D^{\nu}B  u^{\lambda}u_{\nu}f_{-\mu\lambda}\ra+\mathrm{h.c.}$ &280 & $i\la\Bb u^{\mu}{f_{+\mu}}^{\nu}u_{\nu}B\ra$ &460 & $\la\Bb\chip u^{\mu}\ra\la u^{\nu}D_{\mu\nu}B \ra+\mathrm{h.c.}$  \\
101 & $\la\Bb\gamf \gamma^{\mu}D^{\nu}B  u^{\lambda}u_{\lambda}f_{-\mu\nu}\ra+\mathrm{h.c.}$ &281 & $\la\Bb\sigma^{\mu\nu}B f_{+\mu\nu}u^{\lambda}u_{\lambda}\ra+\mathrm{h.c.}$ &461 & $\la\Bb u^{\mu}{D_{\mu}}^{\nu}B  u_{\nu}\ra\la\chip\ra$  \\
102 & $\la\Bb\gamf \gamma^{\mu}D^{\nu}B  u_{\mu}{f_{-\nu}}^{\lambda}u_{\lambda}\ra+\mathrm{h.c.}$ &282 & $\la\Bb\sigma^{\mu\nu}B{f_{+\mu}}^{\lambda}u_{\nu}u_{\lambda}\ra+\mathrm{h.c.}$ &462 & $\la\Bb\gamf \gamma^{\mu}D^{\nu}B  f_{-\mu\nu}\chip\ra+\mathrm{h.c.}$  \\
103 & $\la\Bb\gamf \gamma^{\mu}D^{\nu}B  u_{\nu}{f_{-\mu}}^{\lambda}u_{\lambda}\ra+\mathrm{h.c.}$ &283 & $\la\Bb\sigma^{\mu\nu}B{f_{+\mu}}^{\lambda}u_{\lambda}u_{\nu}\ra+\mathrm{h.c.}$ &463 & $\la\Bb f_{-}^{\mu\nu}\chip\gamf \gamma_{\mu}D_{\nu}B \ra+\mathrm{h.c.}$  \\
104 & $\la\Bb u^{\mu}\gamf \gamma_{\mu}D^{\nu}B  u^{\lambda}f_{-\nu\lambda}\ra+\mathrm{h.c.}$ &284 & $\la\Bb\sigma^{\mu\nu}B u_{\mu}{f_{+\nu}}^{\lambda}u_{\lambda}\ra+\mathrm{h.c.}$ &464 & $\la\Bb\gamf \gamma^{\mu}D^{\nu}B  h_{\mu\nu}\chip\ra+\mathrm{h.c.}$  \\
105 & $\la\Bb u^{\mu}\gamf \gamma^{\nu}D_{\mu}B  u^{\lambda}f_{-\nu\lambda}\ra+\mathrm{h.c.}$ &285 & $\la\Bb\sigma^{\mu\nu}B u^{\lambda}f_{+\mu\nu}u_{\lambda}\ra$ &465 & $\la\Bb h^{\mu\nu}\chip\gamf \gamma_{\mu}D_{\nu}B \ra+\mathrm{h.c.}$  \\
106 & $\la\Bb u^{\mu}\gamf \gamma^{\nu}D^{\lambda}B  u_{\mu}f_{-\nu\lambda}\ra+\mathrm{h.c.}$ &286 & $\la\Bb f_{+}^{\mu\nu}\sigma_{\mu\nu}B u^{\lambda}u_{\lambda}\ra$ &466 & $\la\Bb\gamf \gamma^{\mu}D^{\nu}B  u_{\nu}\chi_{+\mu}\ra+\mathrm{h.c.}$  \\
107 & $\la\Bb u^{\mu}\gamf \gamma^{\nu}D^{\lambda}B  u_{\nu}f_{-\mu\lambda}\ra+\mathrm{h.c.}$ &287 & $\la\Bb f_{+}^{\mu\nu}{\sigma_{\mu}}^{\lambda}B u_{\nu}u_{\lambda}\ra+\mathrm{h.c.}$ &467 & $\la\Bb u^{\mu}\chi_{+}^{\nu}\gamf \gamma_{\nu}D_{\mu}B \ra+\mathrm{h.c.}$  \\
108 & $\la\Bb u^{\mu}\gamf \gamma^{\nu}D^{\lambda}B  u_{\lambda}f_{-\mu\nu}\ra+\mathrm{h.c.}$ &288 & $\la\Bb u^{\mu}{\sigma_{\mu}}^{\nu}B{f_{+\nu}}^{\lambda}u_{\lambda}\ra+\mathrm{h.c.}$ &468 & $\la\Bb\chip\ra\la f_{-}^{\mu\nu}\gamf \gamma_{\mu}D_{\nu}B \ra+\mathrm{h.c.}$  \\
109 & $\la\Bb f_{-}^{\mu\nu}\gamf \gamma_{\mu}D^{\lambda}B  u_{\nu}u_{\lambda}\ra+\mathrm{h.c.}$ &289 & $\la\Bb u^{\mu}\sigma^{\nu\lambda}B f_{+\mu\nu}u_{\lambda}\ra+\mathrm{h.c.}$ &469 & $\la\Bb\chip\ra\la h^{\mu\nu}\gamf \gamma_{\mu}D_{\nu}B \ra+\mathrm{h.c.}$  \\
110 & $\la\Bb f_{-}^{\mu\nu}\gamf \gamma^{\lambda}D_{\mu}B  u_{\nu}u_{\lambda}\ra+\mathrm{h.c.}$ &290 & $\la\Bb u^{\mu}\sigma^{\nu\lambda}B f_{+\nu\lambda}u_{\mu}\ra+\mathrm{h.c.}$ &470 & $\la\Bb\chi_{+}^{\mu}\ra\la u^{\nu}\gamf \gamma_{\mu}D_{\nu}B \ra+\mathrm{h.c.}$  \\
111 & $\la\Bb u^{\mu}u^{\nu}\gamf \gamma_{\mu}D^{\lambda}B  f_{-\nu\lambda}\ra+\mathrm{h.c.}$ &291 & $\la\Bb f_{+}^{\mu\nu}u_{\mu}{\sigma_{\nu}}^{\lambda}B u_{\lambda}\ra+\mathrm{h.c.}$ &471 & $\la\Bb B{\chi_{+}^{\mu}}_{\mu}\ra$  \\
112 & $\la\Bb u^{\mu}u^{\nu}\gamf \gamma^{\lambda}D_{\mu}B  f_{-\nu\lambda}\ra+\mathrm{h.c.}$ &292 & $\la\Bb f_{+}^{\mu\nu}u^{\lambda}\sigma_{\mu\nu}B u_{\lambda}\ra+\mathrm{h.c.}$ &472 & $\la\Bb{\chi_{+}^{\mu}}_{\mu}B\ra$  \\
113 & $\la\Bb u^{\mu}{f_{-\mu}}^{\nu}\gamf \gamma_{\nu}D^{\lambda}B  u_{\lambda}\ra+\mathrm{h.c.}$ &293 & $\la\Bb f_{+}^{\mu\nu}u^{\lambda}\sigma_{\mu\lambda}B u_{\nu}\ra+\mathrm{h.c.}$ &473 & $\la\Bb B\ra\la{\chi_{+}^{\mu}}_{\mu}\ra$  \\
114 & $\la\Bb u^{\mu}{f_{-\mu}}^{\nu}\gamf \gamma^{\lambda}D_{\nu}B  u_{\lambda}\ra+\mathrm{h.c.}$ &294 & $\la\Bb u^{\mu}u_{\mu}\sigma^{\nu\lambda}B f_{+\nu\lambda}\ra$ &474 & $\la\Bb\sigma^{\mu\nu}B f_{+\mu\nu}\chip\ra+\mathrm{h.c.}$  \\
115 & $\la\Bb u^{\mu}f_{-}^{\nu\lambda}\gamf \gamma_{\mu}D_{\nu}B  u_{\lambda}\ra+\mathrm{h.c.}$ &295 & $\la\Bb u^{\mu}u^{\nu}{\sigma_{\mu}}^{\lambda}B f_{+\nu\lambda}\ra+\mathrm{h.c.}$ &475 & $\la\Bb f_{+}^{\mu\nu}\sigma_{\mu\nu}B\chip\ra$  \\
116 & $\la\Bb u^{\mu}f_{-}^{\nu\lambda}\gamf \gamma_{\nu}D_{\mu}B  u_{\lambda}\ra+\mathrm{h.c.}$ &296 & $\la\Bb f_{+}^{\mu\nu}u_{\mu}u^{\lambda}\sigma_{\nu\lambda}B\ra+\mathrm{h.c.}$ &476 & $\la\Bb\chip\sigma^{\mu\nu}B f_{+\mu\nu}\ra$  \\
117 & $\la\Bb u^{\mu}f_{-}^{\nu\lambda}\gamf \gamma_{\nu}D_{\lambda}B  u_{\mu}\ra+\mathrm{h.c.}$ &297 & $\la\Bb f_{+}^{\mu\nu}u^{\lambda}u_{\mu}\sigma_{\nu\lambda}B\ra+\mathrm{h.c.}$ &477 & $\la\Bb f_{+}^{\mu\nu}\chip\sigma_{\mu\nu}B\ra+\mathrm{h.c.}$  \\
118 & $\la\Bb u^{\mu}u_{\mu}f_{-}^{\nu\lambda}\gamf \gamma_{\nu}D_{\lambda}B \ra+\mathrm{h.c.}$ &298 & $\la\Bb f_{+}^{\mu\nu}u^{\lambda}u_{\lambda}\sigma_{\mu\nu}B\ra+\mathrm{h.c.}$ &478 & $\la\Bb\sigma^{\mu\nu}B\ra\la f_{+\mu\nu}\chip\ra$  \\
119 & $\la\Bb u^{\mu}u^{\nu}{f_{-\mu}}^{\lambda}\gamf \gamma_{\nu}D_{\lambda}B \ra+\mathrm{h.c.}$ &299 & $\la\Bb u^{\mu}{f_{+\mu}}^{\nu}u^{\lambda}\sigma_{\nu\lambda}B\ra+\mathrm{h.c.}$ &479 & $\la\Bb\sigma^{\mu\nu}B f_{+\mu\nu}\ra\la\chip\ra$  \\
120 & $\la\Bb u^{\mu}u^{\nu}{f_{-\mu}}^{\lambda}\gamf \gamma_{\lambda}D_{\nu}B \ra+\mathrm{h.c.}$ &300 & $\la\Bb u^{\mu}f_{+}^{\nu\lambda}u_{\mu}\sigma_{\nu\lambda}B\ra$ &480 & $\la\Bb\chip\ra\la f_{+}^{\mu\nu}\sigma_{\mu\nu}B\ra+\mathrm{h.c.}$  \\
121 & $\la\Bb u^{\mu}u^{\nu}{f_{-\nu}}^{\lambda}\gamf \gamma_{\mu}D_{\lambda}B \ra+\mathrm{h.c.}$ &301 & $i\la\Bb D^{\mu\nu}B {f_{+\mu}}^{\lambda}u_{\nu}u_{\lambda}\ra+\mathrm{h.c.}$ &481 & $\la\Bb B\chip^2\ra$  \\
122 & $\la\Bb u^{\mu}u^{\nu}{f_{-\nu}}^{\lambda}\gamf \gamma_{\lambda}D_{\mu}B \ra+\mathrm{h.c.}$ &302 & $i\la\Bb D^{\mu\nu}B {f_{+\mu}}^{\lambda}u_{\lambda}u_{\nu}\ra+\mathrm{h.c.}$ &482 & $\la\Bb\chip B\chip\ra$  \\
123 & $\la\Bb u^{\mu}{f_{-\mu}}^{\nu}u^{\lambda}\gamf \gamma_{\nu}D_{\lambda}B \ra+\mathrm{h.c.}$ &303 & $i\la\Bb D^{\mu\nu}B  u_{\mu}{f_{+\nu}}^{\lambda}u_{\lambda}\ra+\mathrm{h.c.}$ &483 & $\la\Bb\chip^2 B\ra$  \\
124 & $\la\Bb u^{\mu}{f_{-\mu}}^{\nu}u^{\lambda}\gamf \gamma_{\lambda}D_{\nu}B \ra+\mathrm{h.c.}$ &304 & $i\la\Bb f_{+}^{\mu\nu}{D_{\mu}}^{\lambda}B  u_{\nu}u_{\lambda}\ra+\mathrm{h.c.}$ &484 & $\la\Bb B\ra\la\chip^2\ra$  \\
125 & $\la\Bb\gamf \gamma^{\mu}D^{\nu}B  u_{\mu}u^{\lambda}h_{\nu\lambda}\ra+\mathrm{h.c.}$ &305 & $i\la\Bb u^{\mu}{D_{\mu}}^{\nu}B {f_{+\nu}}^{\lambda}u_{\lambda}\ra+\mathrm{h.c.}$ &485 & $\la\Bb B\chip\ra\la\chip\ra$  \\
126 & $\la\Bb\gamf \gamma^{\mu}D^{\nu}B  u_{\nu}u^{\lambda}h_{\mu\lambda}\ra+\mathrm{h.c.}$ &306 & $i\la\Bb u^{\mu}D^{\nu\lambda}B  f_{+\mu\nu}u_{\lambda}\ra+\mathrm{h.c.}$ &486 & $\la\Bb\chip\ra\la\chip B\ra$  \\
127 & $\la\Bb\gamf \gamma^{\mu}D^{\nu}B  u^{\lambda}u_{\mu}h_{\nu\lambda}\ra+\mathrm{h.c.}$ &307 & $i\la\Bb f_{+}^{\mu\nu}u_{\mu}{D_{\nu}}^{\lambda}B  u_{\lambda}\ra+\mathrm{h.c.}$ &487 & $\la\Bb\chip B\ra\la\chip\ra$  \\
128 & $\la\Bb\gamf \gamma^{\mu}D^{\nu}B  u^{\lambda}u_{\nu}h_{\mu\lambda}\ra+\mathrm{h.c.}$ &308 & $i\la\Bb f_{+}^{\mu\nu}u^{\lambda}D_{\mu\lambda}B  u_{\nu}\ra+\mathrm{h.c.}$ &488 & $i\la\Bb\gamf \gamma^{\mu}D^{\nu}B  u_{\mu}u_{\nu}\chim\ra+\mathrm{h.c.}$  \\
129 & $\la\Bb\gamf \gamma^{\mu}D^{\nu}B  u^{\lambda}u_{\lambda}h_{\mu\nu}\ra+\mathrm{h.c.}$ &309 & $i\la\Bb u^{\mu}u^{\nu}{D_{\mu}}^{\lambda}B  f_{+\nu\lambda}\ra+\mathrm{h.c.}$ &489 & $i\la\Bb\gamf \gamma^{\mu}D^{\nu}B  u_{\nu}u_{\mu}\chim\ra+\mathrm{h.c.}$  \\
130 & $\la\Bb\gamf \gamma^{\mu}D^{\nu}B  u_{\nu}{h_{\mu}}^{\lambda}u_{\lambda}\ra+\mathrm{h.c.}$ &310 & $i\la\Bb f_{+}^{\mu\nu}u_{\mu}u^{\lambda}D_{\nu\lambda}B \ra+\mathrm{h.c.}$ &490 & $i\la\Bb\gamf \gamma^{\mu}D^{\nu}B  u_{\mu}\chim u_{\nu}\ra+\mathrm{h.c.}$  \\
131 & $\la\Bb u^{\mu}\gamf \gamma_{\mu}D^{\nu}B  u^{\lambda}h_{\nu\lambda}\ra+\mathrm{h.c.}$ &311 & $i\la\Bb f_{+}^{\mu\nu}u^{\lambda}u_{\mu}D_{\nu\lambda}B \ra+\mathrm{h.c.}$ &491 & $i\la\Bb u^{\mu}\gamf \gamma_{\mu}D^{\nu}B  u_{\nu}\chim\ra+\mathrm{h.c.}$  \\
132 & $\la\Bb u^{\mu}\gamf \gamma^{\nu}D_{\mu}B  u^{\lambda}h_{\nu\lambda}\ra+\mathrm{h.c.}$ &312 & $i\la\Bb u^{\mu}{f_{+\mu}}^{\nu}u^{\lambda}D_{\nu\lambda}B \ra+\mathrm{h.c.}$ &492 & $i\la\Bb u^{\mu}\gamf \gamma^{\nu}D_{\mu}B  u_{\nu}\chim\ra+\mathrm{h.c.}$  \\
133 & $\la\Bb u^{\mu}\gamf \gamma^{\nu}D^{\lambda}B  u_{\mu}h_{\nu\lambda}\ra+\mathrm{h.c.}$ &313 & $\la\Bb\sigma^{\mu\nu}D^{\lambda\rho}B  f_{+\mu\nu}u_{\lambda}u_{\rho}\ra+\mathrm{h.c.}$ &493 & $i\la\Bb\chim\gamf \gamma^{\mu}D^{\nu}B  u_{\mu}u_{\nu}\ra+\mathrm{h.c.}$  \\
134 & $\la\Bb u^{\mu}\gamf \gamma^{\nu}D^{\lambda}B  u_{\nu}h_{\mu\lambda}\ra+\mathrm{h.c.}$ &314 & $\la\Bb\sigma^{\mu\nu}D^{\lambda\rho}B  f_{+\mu\lambda}u_{\nu}u_{\rho}\ra+\mathrm{h.c.}$ &494 & $i\la\Bb u^{\mu}u^{\nu}\gamf \gamma_{\mu}D_{\nu}B \chim\ra+\mathrm{h.c.}$  \\
135 & $\la\Bb u^{\mu}\gamf \gamma^{\nu}D^{\lambda}B  u_{\lambda}h_{\mu\nu}\ra+\mathrm{h.c.}$ &315 & $\la\Bb\sigma^{\mu\nu}D^{\lambda\rho}B  f_{+\mu\lambda}u_{\rho}u_{\nu}\ra+\mathrm{h.c.}$ &495 & $i\la\Bb u^{\mu}\chim\gamf \gamma_{\mu}D^{\nu}B  u_{\nu}\ra+\mathrm{h.c.}$  \\
136 & $\la\Bb h^{\mu\nu}\gamf \gamma_{\mu}D^{\lambda}B  u_{\nu}u_{\lambda}\ra+\mathrm{h.c.}$ &316 & $\la\Bb\sigma^{\mu\nu}D^{\lambda\rho}B  u_{\mu}f_{+\nu\lambda}u_{\rho}\ra+\mathrm{h.c.}$ &496 & $i\la\Bb u^{\mu}\chim\gamf \gamma^{\nu}D_{\mu}B  u_{\nu}\ra+\mathrm{h.c.}$  \\
137 & $\la\Bb u^{\mu}u^{\nu}\gamf \gamma_{\mu}D^{\lambda}B  h_{\nu\lambda}\ra+\mathrm{h.c.}$ &317 & $\la\Bb\sigma^{\mu\nu}D^{\lambda\rho}B  u_{\lambda}f_{+\mu\nu}u_{\rho}\ra$ &497 & $i\la\Bb u^{\mu}u^{\nu}\chim\gamf \gamma_{\mu}D_{\nu}B \ra+\mathrm{h.c.}$  \\
138 & $\la\Bb u^{\mu}u^{\nu}\gamf \gamma^{\lambda}D_{\mu}B  h_{\nu\lambda}\ra+\mathrm{h.c.}$ &318 & $\la\Bb f_{+}^{\mu\nu}\sigma_{\mu\nu}D^{\lambda\rho}B  u_{\lambda}u_{\rho}\ra$ &498 & $i\la\Bb u^{\mu}u^{\nu}\chim\gamf \gamma_{\nu}D_{\mu}B \ra+\mathrm{h.c.}$  \\
139 & $\la\Bb u^{\mu}{h_{\mu}}^{\nu}\gamf \gamma_{\nu}D^{\lambda}B  u_{\lambda}\ra+\mathrm{h.c.}$ &319 & $\la\Bb f_{+}^{\mu\nu}{\sigma_{\mu}}^{\lambda}{D_{\nu}}^{\rho}B  u_{\lambda}u_{\rho}\ra+\mathrm{h.c.}$ &499 & $i\la\Bb u^{\mu}\chim u^{\nu}\gamf \gamma_{\mu}D_{\nu}B \ra+\mathrm{h.c.}$  \\
140 & $\la\Bb u^{\mu}{h_{\mu}}^{\nu}\gamf \gamma^{\lambda}D_{\nu}B  u_{\lambda}\ra+\mathrm{h.c.}$ &320 & $\la\Bb u^{\mu}{\sigma_{\mu}}^{\nu}D^{\lambda\rho}B  f_{+\nu\lambda}u_{\rho}\ra+\mathrm{h.c.}$ &500 & $i\la\Bb\gamf \gamma^{\mu}D^{\nu}B \ra\la u_{\mu}u_{\nu}\chim\ra+\mathrm{h.c.}$  \\
141 & $\la\Bb u^{\mu}h^{\nu\lambda}\gamf \gamma_{\mu}D_{\nu}B  u_{\lambda}\ra+\mathrm{h.c.}$ &321 & $\la\Bb u^{\mu}\sigma^{\nu\lambda}{D_{\mu}}^{\rho}B  f_{+\nu\lambda}u_{\rho}\ra+\mathrm{h.c.}$ &501 & $i\la\Bb\gamf \gamma^{\mu}D^{\nu}B  u_{\mu}u_{\nu}\ra\la\chim\ra+\mathrm{h.c.}$  \\
142 & $\la\Bb u^{\mu}h^{\nu\lambda}\gamf \gamma_{\nu}D_{\mu}B  u_{\lambda}\ra+\mathrm{h.c.}$ &322 & $\la\Bb u^{\mu}\sigma^{\nu\lambda}{D_{\mu}}^{\rho}B  f_{+\nu\rho}u_{\lambda}\ra+\mathrm{h.c.}$ &502 & $i\la\Bb u^{\mu}\chim\ra\la u^{\nu}\gamf \gamma_{\nu}D_{\mu}B \ra+\mathrm{h.c.}$  \\
143 & $\la\Bb u^{\mu}u_{\mu}h^{\nu\lambda}\gamf \gamma_{\nu}D_{\lambda}B \ra+\mathrm{h.c.}$ &323 & $\la\Bb f_{+}^{\mu\nu}u^{\lambda}\sigma_{\mu\nu}{D_{\lambda}}^{\rho}B  u_{\rho}\ra+\mathrm{h.c.}$ &503 & $i\la\Bb\chim\ra\la u^{\mu}u^{\nu}\gamf \gamma_{\mu}D_{\nu}B \ra+\mathrm{h.c.}$  \\
144 & $\la\Bb u^{\mu}u^{\nu}{h_{\mu}}^{\lambda}\gamf \gamma_{\nu}D_{\lambda}B \ra+\mathrm{h.c.}$ &324 & $\la\Bb f_{+}^{\mu\nu}u^{\lambda}\sigma_{\mu\lambda}{D_{\nu}}^{\rho}B  u_{\rho}\ra+\mathrm{h.c.}$ &504 & $i\la\Bb u^{\mu}\chim\ra\la u^{\nu}\gamf \gamma_{\mu}D_{\nu}B \ra+\mathrm{h.c.}$  \\
145 & $\la\Bb u^{\mu}u^{\nu}{h_{\mu}}^{\lambda}\gamf \gamma_{\lambda}D_{\nu}B \ra+\mathrm{h.c.}$ &325 & $\la\Bb f_{+}^{\mu\nu}u^{\lambda}{\sigma_{\mu}}^{\rho}D_{\nu\lambda}B  u_{\rho}\ra+\mathrm{h.c.}$ &505 & $\la\Bb\sigma^{\mu\nu}B f_{-\mu\nu}\chim\ra+\mathrm{h.c.}$  \\
146 & $\la\Bb u^{\mu}u^{\nu}{h_{\nu}}^{\lambda}\gamf \gamma_{\mu}D_{\lambda}B \ra+\mathrm{h.c.}$ &326 & $\la\Bb u^{\mu}u^{\nu}{\sigma_{\mu}}^{\lambda}{D_{\nu}}^{\rho}B  f_{+\lambda\rho}\ra+\mathrm{h.c.}$ &506 & $\la\Bb f_{-}^{\mu\nu}\chim\sigma_{\mu\nu}B\ra+\mathrm{h.c.}$  \\
147 & $\la\Bb u^{\mu}u^{\nu}{h_{\nu}}^{\lambda}\gamf \gamma_{\lambda}D_{\mu}B \ra+\mathrm{h.c.}$ &327 & $\la\Bb u^{\mu}u^{\nu}\sigma^{\lambda\rho}D_{\mu\nu}B  f_{+\lambda\rho}\ra$ &507 & $i\la\Bb D^{\mu\nu}B  h_{\mu\nu}\chim\ra+\mathrm{h.c.}$  \\
148 & $\la\Bb u^{\mu}{h_{\mu}}^{\nu}u^{\lambda}\gamf \gamma_{\nu}D_{\lambda}B \ra+\mathrm{h.c.}$ &328 & $\la\Bb f_{+}^{\mu\nu}u^{\lambda}u^{\rho}\sigma_{\mu\nu}D_{\lambda\rho}B \ra+\mathrm{h.c.}$ &508 & $i\la\Bb h^{\mu\nu}D_{\mu\nu}B \chim\ra$  \\
149 & $\epsilon^{\mu\nu\lambda\rho}\la\Bb{D_{\mu}}^{\sigma}B  u_{\nu}u_{\lambda}f_{-\rho\sigma}\ra+\mathrm{h.c.}$ &329 & $\la\Bb f_{+}^{\mu\nu}u^{\lambda}u^{\rho}\sigma_{\mu\lambda}D_{\nu\rho}B \ra+\mathrm{h.c.}$ &509 & $i\la\Bb\chim D^{\mu\nu}B  h_{\mu\nu}\ra$  \\
150 & $\epsilon^{\mu\nu\lambda\rho}\la\Bb{D_{\mu}}^{\sigma}B  u_{\nu}u_{\sigma}f_{-\lambda\rho}\ra+\mathrm{h.c.}$ &330 & $\la\Bb f_{+}^{\mu\nu}u^{\lambda}u^{\rho}\sigma_{\mu\rho}D_{\nu\lambda}B \ra+\mathrm{h.c.}$ &510 & $i\la\Bb h^{\mu\nu}\chim D_{\mu\nu}B \ra+\mathrm{h.c.}$  \\
151 & $\epsilon^{\mu\nu\lambda\rho}\la\Bb{D_{\mu}}^{\sigma}B  u_{\sigma}u_{\nu}f_{-\lambda\rho}\ra+\mathrm{h.c.}$ &331 & $\la\Bb u^{\mu}f_{+}^{\nu\lambda}u^{\rho}\sigma_{\mu\nu}D_{\lambda\rho}B \ra+\mathrm{h.c.}$ &511 & $i\la\Bb B u^{\mu}\chi_{-\mu}\ra+\mathrm{h.c.}$  \\
152 & $\epsilon^{\mu\nu\lambda\rho}\la\Bb{D_{\mu}}^{\sigma}B  u_{\nu}f_{-\lambda\rho}u_{\sigma}\ra+\mathrm{h.c.}$ &332 & $\la\Bb u^{\mu}f_{+}^{\nu\lambda}u^{\rho}\sigma_{\nu\lambda}D_{\mu\rho}B \ra$ &512 & $i\la\Bb u^{\mu}B\chi_{-\mu}\ra$  \\
153 & $\epsilon^{\mu\nu\lambda\rho}\la\Bb u_{\mu}{D_{\nu}}^{\sigma}B  u_{\lambda}f_{-\rho\sigma}\ra+\mathrm{h.c.}$ &333 & $i\la\Bb B\ra\la f_{+}^{\mu\nu}u_{\mu}u_{\nu}\ra$ &513 & $i\la\Bb\chi_{-}^{\mu}B u_{\mu}\ra$  \\
154 & $\epsilon^{\mu\nu\lambda\rho}\la\Bb u_{\mu}{D_{\nu}}^{\sigma}B  u_{\sigma}f_{-\lambda\rho}\ra+\mathrm{h.c.}$ &334 & $\la\Bb\sigma^{\mu\nu}B\ra\la f_{+\mu\nu}u^{\lambda}u_{\lambda}\ra$ &514 & $i\la\Bb u^{\mu}\chi_{-\mu}B\ra+\mathrm{h.c.}$  \\
155 & $\epsilon^{\mu\nu\lambda\rho}\la\Bb u^{\sigma}D_{\mu\sigma}B  u_{\nu}f_{-\lambda\rho}\ra+\mathrm{h.c.}$ &335 & $\la\Bb\sigma^{\mu\nu}B f_{+\mu\nu}\ra\la u^{\lambda}u_{\lambda}\ra$ &515 & $\la\Bb\sigma^{\mu\nu}B u_{\mu}\chi_{-\nu}\ra+\mathrm{h.c.}$  \\
156 & $\epsilon^{\mu\nu\lambda\rho}\la\Bb f_{-\mu\nu}{D_{\lambda}}^{\sigma}B  u_{\rho}u_{\sigma}\ra+\mathrm{h.c.}$ &336 & $\la\Bb u^{\mu}u_{\mu}\ra\la f_{+}^{\nu\lambda}\sigma_{\nu\lambda}B\ra+\mathrm{h.c.}$ &516 & $\la\Bb u^{\mu}\chi_{-}^{\nu}\sigma_{\mu\nu}B\ra+\mathrm{h.c.}$  \\
157 & $\epsilon^{\mu\nu\lambda\rho}\la\Bb u_{\mu}u^{\sigma}D_{\nu\sigma}B  f_{-\lambda\rho}\ra+\mathrm{h.c.}$ &337 & $\la\Bb\sigma^{\mu\nu}B\ra\la{f_{+\mu}}^{\lambda}u_{\nu}u_{\lambda}\ra+\mathrm{h.c.}$ &517 & $\la\Bb\chim\ra\la f_{-}^{\mu\nu}\sigma_{\mu\nu}B\ra+\mathrm{h.c.}$  \\
158 & $\epsilon^{\mu\nu\lambda\rho}\la\Bb u_{\mu}f_{-\nu\lambda}{D_{\rho}}^{\sigma}B  u_{\sigma}\ra+\mathrm{h.c.}$ &338 & $\la\Bb\sigma^{\mu\nu}B{f_{+\mu}}^{\lambda}\ra\la u_{\nu}u_{\lambda}\ra$ &518 & $i\la\Bb D^{\mu\nu}B \ra\la h_{\mu\nu}\chim\ra$  \\
159 & $\epsilon^{\mu\nu\lambda\rho}\la\Bb u_{\mu}{f_{-\nu}}^{\sigma}D_{\lambda\sigma}B  u_{\rho}\ra+\mathrm{h.c.}$ &339 & $\la\Bb u^{\mu}u^{\nu}\ra\la{f_{+\nu}}^{\lambda}\sigma_{\mu\lambda}B\ra+\mathrm{h.c.}$ &519 & $i\la\Bb D^{\mu\nu}B  h_{\mu\nu}\ra\la\chim\ra$  \\
160 & $\epsilon^{\mu\nu\lambda\rho}\la\Bb u^{\sigma}f_{-\mu\nu}D_{\lambda\sigma}B  u_{\rho}\ra+\mathrm{h.c.}$ &340 & $\la\Bb u^{\mu}\ra\la{f_{+\mu}}^{\nu}u^{\lambda}\sigma_{\nu\lambda}B\ra+\mathrm{h.c.}$ &520 & $i\la\Bb\chim\ra\la h^{\mu\nu}D_{\mu\nu}B \ra+\mathrm{h.c.}$  \\
161 & $\epsilon^{\mu\nu\lambda\rho}\la\Bb u_{\mu}u_{\nu}{f_{-\lambda}}^{\sigma}D_{\rho\sigma}B \ra+\mathrm{h.c.}$ &341 & $\la\Bb u^{\mu}u^{\nu}\ra\la{f_{+\mu}}^{\lambda}\sigma_{\nu\lambda}B\ra+\mathrm{h.c.}$ &521 & $i\la\Bb B\ra\la u^{\mu}\chi_{-\mu}\ra$  \\
162 & $\epsilon^{\mu\nu\lambda\rho}\la\Bb u_{\mu}u^{\sigma}f_{-\nu\lambda}D_{\rho\sigma}B \ra+\mathrm{h.c.}$ &342 & $\la\Bb f_{+}^{\mu\nu}u_{\mu}\ra\la u^{\lambda}\sigma_{\nu\lambda}B\ra+\mathrm{h.c.}$ &522 & $i\la\Bb B u^{\mu}\ra\la\chi_{-\mu}\ra$  \\
163 & $\epsilon^{\mu\nu\lambda\rho}\la\Bb u^{\sigma}u_{\mu}f_{-\nu\lambda}D_{\rho\sigma}B \ra+\mathrm{h.c.}$ &343 & $i\la\Bb D^{\mu\nu}B \ra\la{f_{+\mu}}^{\lambda}u_{\nu}u_{\lambda}\ra+\mathrm{h.c.}$ &523 & $i\la\Bb\chi_{-}^{\mu}\ra\la u_{\mu}B\ra+\mathrm{h.c.}$  \\
164 & $\epsilon^{\mu\nu\lambda\rho}\la\Bb u_{\mu}f_{-\nu\lambda}u^{\sigma}D_{\rho\sigma}B \ra+\mathrm{h.c.}$ &344 & $i\la\Bb u^{\mu}u^{\nu}\ra\la{f_{+\nu}}^{\lambda}D_{\mu\lambda}B \ra+\mathrm{h.c.}$ &524 & $\la\Bb\chi_{-}^{\mu}\ra\la u^{\nu}\sigma_{\mu\nu}B\ra+\mathrm{h.c.}$  \\
165 & $\epsilon^{\mu\nu\lambda\rho}\la\Bb{D_{\mu}}^{\sigma}B  u_{\nu}u_{\lambda}h_{\rho\sigma}\ra+\mathrm{h.c.}$ &345 & $\la\Bb\sigma^{\mu\nu}D^{\lambda\rho}B \ra\la f_{+\mu\nu}u_{\lambda}u_{\rho}\ra$ &525 & $\la\Bb\gamf \gamma^{\mu}D^{\nu}B  f_{+\mu\nu}\chim\ra+\mathrm{h.c.}$  \\
166 & $\epsilon^{\mu\nu\lambda\rho}\la\Bb u_{\mu}{D_{\nu}}^{\sigma}B  u_{\lambda}h_{\rho\sigma}\ra+\mathrm{h.c.}$ &346 & $\la\Bb\sigma^{\mu\nu}D^{\lambda\rho}B  f_{+\mu\nu}\ra\la u_{\lambda}u_{\rho}\ra$ &526 & $\la\Bb f_{+}^{\mu\nu}\gamf \gamma_{\mu}D_{\nu}B \chim\ra$  \\
167 & $\epsilon^{\mu\nu\lambda\rho}\la\Bb u_{\mu}{h_{\nu}}^{\sigma}D_{\lambda\sigma}B  u_{\rho}\ra+\mathrm{h.c.}$ &347 & $\la\Bb u^{\mu}u^{\nu}\ra\la f_{+}^{\lambda\rho}\sigma_{\lambda\rho}D_{\mu\nu}B \ra+\mathrm{h.c.}$ &527 & $\la\Bb\chim\gamf \gamma^{\mu}D^{\nu}B  f_{+\mu\nu}\ra$  \\
168 & $\epsilon^{\mu\nu\lambda\rho}\la\Bb u_{\mu}u_{\nu}{h_{\lambda}}^{\sigma}D_{\rho\sigma}B \ra+\mathrm{h.c.}$ &348 & $\la\Bb\sigma^{\mu\nu}D^{\lambda\rho}B \ra\la f_{+\mu\lambda}u_{\nu}u_{\rho}\ra+\mathrm{h.c.}$ &528 & $\la\Bb f_{+}^{\mu\nu}\chim\gamf \gamma_{\mu}D_{\nu}B \ra+\mathrm{h.c.}$  \\
169 & $\la\Bb\gamf \gamma^{\mu}D^{\nu\lambda\rho}B  u_{\nu}u_{\lambda}f_{-\mu\rho}\ra+\mathrm{h.c.}$ &349 & $\la\Bb\sigma^{\mu\nu}D^{\lambda\rho}B  f_{+\mu\lambda}\ra\la u_{\nu}u_{\rho}\ra$ &529 & $\la\Bb\gamf \gamma^{\mu}D^{\nu}B \ra\la f_{+\mu\nu}\chim\ra$  \\
170 & $\la\Bb u^{\mu}\gamf \gamma^{\nu}{D_{\mu}}^{\lambda\rho}B  u_{\lambda}f_{-\nu\rho}\ra+\mathrm{h.c.}$ &350 & $\la\Bb u^{\mu}u^{\nu}\ra\la f_{+}^{\lambda\rho}\sigma_{\mu\lambda}D_{\nu\rho}B \ra+\mathrm{h.c.}$ &530 & $\la\Bb\gamf \gamma^{\mu}D^{\nu}B  f_{+\mu\nu}\ra\la\chim\ra$  \\
171 & $\la\Bb u^{\mu}f_{-}^{\nu\lambda}\gamf \gamma_{\nu}{D_{\mu\lambda}}^{\rho}B  u_{\rho}\ra+\mathrm{h.c.}$ &351 & $\la\Bb u^{\mu}\ra\la f_{+}^{\nu\lambda}u^{\rho}\sigma_{\nu\rho}D_{\mu\lambda}B \ra+\mathrm{h.c.}$ &531 & $\la\Bb\chim\ra\la f_{+}^{\mu\nu}\gamf \gamma_{\mu}D_{\nu}B \ra+\mathrm{h.c.}$  \\
172 & $\la\Bb u^{\mu}u^{\nu}f_{-}^{\lambda\rho}\gamf \gamma_{\lambda}D_{\mu\nu\rho}B \ra+\mathrm{h.c.}$ &352 & $\la\Bb u^{\mu}u^{\nu}\ra\la f_{+}^{\lambda\rho}\sigma_{\nu\lambda}D_{\mu\rho}B \ra+\mathrm{h.c.}$ &532 & $\la\Bb B\chim^2\ra$  \\
173 & $\la\Bb\gamf \gamma^{\mu}D^{\nu\lambda\rho}B  u_{\mu}u_{\nu}h_{\lambda\rho}\ra+\mathrm{h.c.}$ &353 & $\la\Bb f_{+}^{\mu\nu}u^{\lambda}\ra\la u^{\rho}\sigma_{\mu\rho}D_{\nu\lambda}B \ra+\mathrm{h.c.}$ &533 & $\la\Bb\chim B\chim\ra$  \\
174 & $\la\Bb\gamf \gamma^{\mu}D^{\nu\lambda\rho}B  u_{\nu}u_{\mu}h_{\lambda\rho}\ra+\mathrm{h.c.}$ &354 & $i\epsilon^{\mu\nu\lambda\rho}\la\Bb B f_{+\mu\nu}f_{-\lambda\rho}\ra+\mathrm{h.c.}$ &534 & $\la\Bb\chim^2 B\ra$  \\
175 & $\la\Bb\gamf \gamma^{\mu}D^{\nu\lambda\rho}B  u_{\nu}u_{\lambda}h_{\mu\rho}\ra+\mathrm{h.c.}$ &355 & $i\epsilon^{\mu\nu\lambda\rho}\la\Bb f_{+\mu\nu}f_{-\lambda\rho}B\ra+\mathrm{h.c.}$ &535 & $\la\Bb B\chim\ra\la\chim\ra$  \\
176 & $\la\Bb u^{\mu}\gamf \gamma_{\mu}D^{\nu\lambda\rho}B  u_{\nu}h_{\lambda\rho}\ra+\mathrm{h.c.}$ &356 & $i\la\Bb\gamf \gamma^{\mu}D^{\nu}B {f_{+\mu}}^{\lambda}f_{-\nu\lambda}\ra+\mathrm{h.c.}$ &536 & $\la\Bb\chim\ra\la\chim B\ra$  \\
177 & $\la\Bb u^{\mu}\gamf \gamma^{\nu}{D_{\mu}}^{\lambda\rho}B  u_{\nu}h_{\lambda\rho}\ra+\mathrm{h.c.}$ &357 & $i\la\Bb\gamf \gamma^{\mu}D^{\nu}B {f_{+\nu}}^{\lambda}f_{-\mu\lambda}\ra+\mathrm{h.c.}$ &537 & $\la\Bb\chim B\ra\la\chim\ra$  \\
178 & $\la\Bb u^{\mu}\gamf \gamma^{\nu}{D_{\mu}}^{\lambda\rho}B  u_{\lambda}h_{\nu\rho}\ra+\mathrm{h.c.}$ &358 & $i\la\Bb f_{+}^{\mu\nu}\gamf \gamma_{\mu}D^{\lambda}B  f_{-\nu\lambda}\ra$ &538 & $\la\Bb B\ra\la\chi\chi^{\dag}\ra$  \\
179 & $\la\Bb u^{\mu}u^{\nu}\gamf \gamma_{\mu}{D_{\nu}}^{\lambda\rho}B  h_{\lambda\rho}\ra+\mathrm{h.c.}$ &359 & $i\la\Bb f_{+}^{\mu\nu}\gamf \gamma^{\lambda}D_{\mu}B  f_{-\nu\lambda}\ra$ &539 & $\la\Bb B\ra\la F_{R}^{\mu\nu}F_{R\mu\nu}\ra+\mathrm{h.c.}$  \\
180 & $\la\Bb u^{\mu}h^{\nu\lambda}\gamf \gamma_{\mu}{D_{\nu\lambda}}^{\rho}B  u_{\rho}\ra+\mathrm{h.c.}$ &360 & $i\la\Bb f_{-}^{\mu\nu}\gamf \gamma_{\mu}D^{\lambda}B  f_{+\nu\lambda}\ra$ &540 & $\la\Bb D^{\mu\nu}B \ra\la{F_{R\mu}}^{\lambda}F_{R\nu\lambda}\ra+\mathrm{h.c.}$  \\
\hline\hline
\end{longtable}
\bibliography{references}
\end{document}